\documentclass[
superscriptaddress,
showpacs,
amssymb,
10pt,
reprint,
aps,
prd,
longbibliography,
nofootinbib,
floatfix
]{revtex4-2}

\usepackage[T1]{fontenc}
\usepackage[utf8]{inputenc}
\usepackage{times}
\usepackage{microtype}

\usepackage[dvipsnames]{xcolor}

\usepackage{amsmath,amssymb,amsfonts}
\usepackage{mathtools}
\usepackage{bm}
\usepackage{bbm}
\usepackage{mathrsfs}
\usepackage{leftindex}
\usepackage{tensor}
\usepackage{accents}
\usepackage{scalerel}

\usepackage{mhchem}
\usepackage{siunitx}

\usepackage{graphicx}
\usepackage{placeins}
\usepackage{subfigure}
\usepackage[percent]{overpic}
\usepackage{tikz}
\usetikzlibrary{calc}

\usepackage{booktabs}
\usepackage{dcolumn}
\usepackage{enumitem}

\usepackage{soul}
\usepackage[normalem]{ulem}
\usepackage{comment}

\usepackage{orcidlink}
\usepackage{xurl}
\usepackage{hyperref}

\definecolor{darkblue}{rgb}{0,0,0.5}
\definecolor{navy}{RGB}{0,0,150}

\hypersetup{
    bookmarks=true,
    pdftoolbar=true,
    pdfmenubar=true,
    pdffitwindow=true,
    pdfstartview={FitH},
    pdftitle={Constitutive birefringence and critical curves in the rotating Garcia-Diaz black hole},
    pdfauthor={Ariel Guzman, Mohsen Fathi, J.R. Villanueva},
    pdfsubject={Nonlinear electrodynamics and black hole critical curves},
    pdfcreator={Creator},
    pdfproducer={Producer},
    pdfkeywords={nonlinear electrodynamics, birefringence, black hole shadows, Garcia-Diaz solution},
    pdfnewwindow=true,
    colorlinks=true,
    linkcolor=darkblue,
    citecolor=darkblue,
    filecolor=navy,
    urlcolor=darkblue
}

\begin{document}


\title{Constitutive birefringence and critical curves in the rotating Garc\'ia--D\'iaz black hole}

\author{Ariel Guzmán\orcidlink{0009-0008-3844-1203}}
\email{ariel.guzman@estudiantes.uv.cl}
\affiliation{Instituto de F\'{i}sica y Astronom\'{i}a, Facultad de Ciencias, Universidad de Valpara\'{i}so,
Avenida Gran Breta\~{n}a 1111, Valpara\'{i}so, Chile}

\author{Mohsen Fathi\orcidlink{0000-0002-1602-0722}}
\email{mohsen.fathi@ucentral.cl}
\affiliation{Centro de Investigaci\'{o}n en Ciencias del Espacio y F\'{i}sica Te\'{o}rica (CICEF), Universidad Central de Chile, La Serena 1710164, Chile}

\author{J.R. Villanueva\orcidlink{0000-0002-6726-492X}}
\email{jose.villanueva@uv.cl}
\affiliation{Instituto de F\'{i}sica y Astronom\'{i}a, Facultad de Ciencias, Universidad de Valpara\'{i}so,
Avenida Gran Breta\~{n}a 1111, Valpara\'{i}so, Chile}



\begin{abstract}
We study high-frequency electromagnetic propagation in the rotating
Garc\'ia--D\'iaz black hole of Einstein gravity coupled to nonlinear
electrodynamics (NLED).  In this setting, light is not fixed only by the
null cone of the spacetime metric.  The nonlinear electromagnetic field also
acts as an optical medium, and its local constitutive response determines the
physical optical cones.  The rotating Garc\'ia--D\'iaz solution is therefore a
useful laboratory, because the metric, the mixed electromagnetic potentials
and the constitutive structure are known within the same exact branch.
Starting from the mixed potentials, we project the field $F$ and the
excitation $P$ on a principal tetrad.  This gives the aligned scalars $E$,
$B$, $D$ and $H$, from which we reconstruct the regular local constitutive
branch connected with Maxwell theory as the map $(D,B)\mapsto(E,H)$.  We then
insert the corresponding response matrix in the Fresnel characteristic
problem.  At the perturbative order considered here, the Fresnel quartic
factorizes into two quadratic branches, each one defining an effective optical
metric.  Both branches admit a Carter-type separation of the Hamilton--Jacobi
equation and therefore have their own radial and angular potentials, critical
constants and unstable critical families.  After projecting these families on
the celestial sphere of a finite-distance observer, we obtain two critical
contours, $\Gamma_+$ and $\Gamma_-$, which coincide in the Maxwell limit and
split when the nonlinear constitutive response is active.  We quantify this
splitting through the maximum angular separation, the relative diameter shift
and the normalized birefringent width.  Numerical scans in the nonlinear
coupling, spin and observer inclination show that the effect is opened by the
constitutive response, redistributed by rotation and preserved under changes
of local projection inside the perturbative domain.  The result gives a direct
geometrical chain from the local NLED response to a polarization-dependent
critical structure on the observer screen.

\bigskip

\end{abstract}

\maketitle


\section{Introduction}
\label{sec:introduction}

In Maxwell electrodynamics the geometrical-optics limit has a simple meaning:
the wave vector of light is null with respect to the spacetime metric.  In
that case, the optical cone and the gravitational null cone are the same, and
photon propagation can be studied directly from the background geometry
\cite{Synge1960,Wald1984}.  This picture changes in nonlinear electrodynamics
(NLED).  The Born--Infeld theory and the Heisenberg--Euler effective action
already showed that the electromagnetic response can depend on the field
itself \cite{BornInfeld1934,HeisenbergEuler1936,BialynickaBirula1970,
Adler1971,DittrichGies2000,Dunne2004,Shore2003}.  The electromagnetic field
then behaves as a medium, and the physical light cone is fixed not only by the
metric, but also by the constitutive law of the theory.

This is the natural language of the covariant formulation of NLED.  In
Pleba\'nski's approach one works with the field $F_{\mu\nu}$, the excitation
$P_{\mu\nu}$ and the electromagnetic invariants.  The mixed formulation of
Salazar, Garc\'ia and Pleba\'nski is especially useful when the relation
between field and excitation is more basic than a single elementary
Lagrangian written globally as $L(\mathcal F,\mathcal G)$
\cite{Plebanski1970,SalazarGarciaPlebanski1987,SalazarGarciaPlebanski1989}.
In this view, the optical problem is not added to the geometry from outside.
It is part of the same electromagnetic response.  The works of Boillat,
Novello et al., and Obukhov and Rubilar showed that the characteristic
equation of a nonlinear electromagnetic theory can become a Fresnel quartic
and, in many cases, can split into two quadratic cones
\cite{Boillat1970,NovelloEtAl2000,ObukhovRubilar2002,HehlObukhov2003,
Rubilar2002}.  This splitting is the geometrical origin of vacuum
birefringence in NLED.

When NLED is coupled to gravity, the electromagnetic field has two roles.  It
sources the spacetime through the Einstein equations, and it also determines
the optical cone followed by high-frequency electromagnetic perturbations
\cite{AyonBeatoGarcia1998,AyonBeatoGarcia1999,AyonBeatoGarcia2000,
Bronnikov2001,Dymnikova1992,BalartVagenas2014,FanWang2016,
Bronnikov2022}.  Therefore, for an Einstein--NLED black hole, null geodesics
of the background metric are not always the physical photon trajectories.  One
has to use the effective optical geometry selected by the constitutive
response.  This point has become important in studies of regular black holes,
effective metrics, light rings, shadows and perturbations
\cite{dePaulaEtAl2023,Walia2024NEDShadows,UniyalEtAl2024,TangEtAl2023,
DePaulaEtAl2026,Fathi:2025jrk}.  In particular, the recent study of axial
electromagnetic quasinormal modes in a static Pleba\'nski-type NLED black hole
shows that the same nonlinear sector can also leave dynamical imprints
\cite{Fathi:2025jrk}.  The present paper is complementary to that direction:
here we focus on the geometrical-optics and critical-curve side.

Black hole shadows are a natural place to see this difference.  Since the
works of Bardeen and Luminet, the edge of the shadow has been related to
unstable photon orbits and to their projection on the observer screen
\cite{Bardeen1973,Luminet1979}.  This idea was later developed for analytical
shadow constructions, finite-distance observers, deformed compact objects and
non-Kerr geometries \cite{FalckeMeliaAgol2000,Takahashi2004,HiokiMaeda2009,
JohannsenPsaltis2010,AbdujabbarovEtAl2013,
GrenzebachPerlickLammerzahl2014,TsukamotoLiBambi2014,YounsiEtAl2016,
CunhaHerdeiroRadu2017,CunhaHerdeiro2018,VagnozziVisinelli2019,
GrallaHolzWald2019,PerlickTsupko2022}.  The observational relevance of the
subject increased after the Event Horizon Telescope images of M87* and
Sagittarius A* \cite{EHT2019I,EHT2019II,EHT2019III,EHT2019IV,EHT2019V,
EHT2019VI,EHT2021VII,EHT2022SgrAI,EHT2022SgrAII,EHT2022SgrAIII,
EHT2022SgrAIV,EHT2022SgrAV,EHT2022SgrAVI}.  Also, the study of photon rings
and higher-order lensing bands has made clear that small changes in the
critical structure can be important in ideal images
\cite{JohnsonEtAl2020,GrallaLupsascaMarrone2020,HimwichEtAl2020,
WielgusEtAl2020,LockhartGralla2022,PaugnatEtAl2022,
KocherlakotaEtAl2024,LupsascaEtAl2024BHEX}.

The present work sits at this intersection: effective optical geometry in
NLED and critical curves of rotating black holes.  For static and spherical
solutions, the optical problem can often be reduced to an effective light-ring
radius.  For a rotating spacetime the problem is richer.  The shadow is
controlled by critical families, impact parameters, angular motion and a local
tetrad projection.  In Kerr and Kerr--Newman geometries, this structure is
tractable because the Hamilton--Jacobi equation separates, as shown by Carter
\cite{Carter1968HJ,Carter1968Global,Chandrasekhar1983,WalkerPenrose1970,
Kinnersley1969}.  This separability is related to hidden symmetries, Killing
tensors and principal tensor structures
\cite{BenentiFrancaviglia1979,Floyd1973,FrolovKrtousKubiznak2017,
KubiznakFrolov2007}.  It is then natural to ask what remains of this
structure when the photon does not follow the spacetime metric itself, but an
optical metric produced by NLED.  We keep contact with standard treatments of
spherical photon motion, Kerr lensing bands and nonequatorial critical
structures \cite{Mino2003,FujitaHikida2009,DattaMukherjee2021,
CunhaHerdeiroRadu2018NoZ2}.

The exact rotating Garc\'ia--D\'iaz solution provides a useful laboratory for
this question.  It is an exact stationary and axially symmetric solution of
Einstein gravity coupled to NLED, with mass, rotation, electric and magnetic
charges, and electrodynamic parameters controlling the nonlinear branch
\cite{GarciaDiaz2021,GarciaDiaz2022Adsds}.  Its Maxwell limit is the dyonic
Kerr--Newman solution \cite{NewmanEtAl1965,NewmanJanis1965}.  For nonzero
nonlinear coupling, both the metric function and the electromagnetic
potentials are modified.  Ay\'on--Beato showed that the electrodynamics behind
this solution is naturally recovered in the mixed formulation, even when the
supporting Lagrangian cannot be written globally in elementary form
\cite{AyonBeato2024}.  Thus, the optical problem should not be treated as a
phenomenological deformation added by hand.  It has to be obtained from the
constitutive response of the exact solution.  This also connects the present
work with studies of rotating regular black holes and nonlinear
electromagnetic sources \cite{Kerr1963,Reissner1916,Nordstrom1918}.  We also
keep contact with Newman--Janis-type constructions and related rotating NLED
models \cite{BambiModesto2013,AzregAinou2014,NevesSaa2014,
ToshmatovStuchlikAhmedov2017,Hendi2012,RodriguesSilva2018}.

In this paper we follow the optical chain step by step.  First, we write the
Garc\'ia--D\'iaz metric and its two mixed potentials in a principal tetrad.
This gives an aligned form for the field and excitation, described by four
scalars $E$, $B$, $D$ and $H$.  Second, we use these scalars to reconstruct
the local constitutive branch connected with the Maxwell limit, written as
$(D,B)\mapsto(E,H)$.  Third, we compute the response matrix and insert it in
the Fresnel construction.  Finally, we study the Hamilton--Jacobi problem,
build the branch-dependent critical families and project them on the screen
of a finite-distance observer.

The main message is simple.  The nonlinear electromagnetic structure does not
only modify the background metric.  It also changes the characteristic cones
that control high-frequency electromagnetic propagation.  For the rotating
Garc\'ia--D\'iaz branch studied here, the Fresnel quartic splits into two
optical roots.  At the perturbative order used in this work, both roots admit
a Carter-type separation.  Each optical branch therefore has its own radial
and angular potentials, its own critical constants and its own projected
critical contour.  The birefringent signal appears as a doubling of the
critical structure: the two curves collapse to one curve in the Maxwell
limit, but separate when the nonlinear response is active.

The paper is organized as follows.  In Sec.~\ref{sec:background} we present
the Garc\'ia--D\'iaz metric, the mixed potentials, the principal tetrad and
the aligned electromagnetic scalars.  In Sec.~\ref{sec:constitutive} we
reconstruct the local constitutive branch and obtain its response matrix.  In
Sec.~\ref{sec:opticalmetrics} we derive the Fresnel cones and the optical
metrics.  In Sec.~\ref{sec:HJ} we discuss the Hamilton--Jacobi equations and
show that both optical branches admit a Carter-type separation at the order
considered.  In Sec.~\ref{sec:birefringent_critical_curves} we construct the
two branch-dependent critical families, project them on a finite-distance
observer screen and introduce geometrical diagnostics of the splitting.  In
Sec.~\ref{sec:discussion} we discuss the physical meaning, scope and
observational relevance of the result.  We close in Sec.~\ref{sec:conclusions}.
Numerical details, parameter scans and reproducibility information are
collected in Appendix~\ref{app:numerical_scans}.


\section{Garc\'ia--D\'iaz background and aligned fields}
\label{sec:background}

We first fix the exact background that will be used in the rest of the paper.
This includes the metric, the mixed electromagnetic potentials, the principal
tetrad and the simple algebraic identities followed by the aligned field.  The
purpose of this section is only to prepare the geometrical and electromagnetic
data.  The constitutive interpretation will be made in the next section, where
the relation between field and excitation is read directly from these exact
quantities.

We work in units $G=c=1$, with signature
\begin{equation}
(-,+,+,+),
\label{eq:signature_convention}
\end{equation}
and in an exterior region where the principal tetrad used below is regular.

\subsection{Metric, parameters, and asymptotic structure}
\label{subsec:metric_parameters_finite_observer}

We adopt as background the rotating García--Díaz solution in
Boyer--Lindquist-type coordinates \cite{GarciaDiaz2021,AyonBeato2024},
\begin{equation}
\begin{aligned}
ds^2={}&
\Sigma\,d\theta^2
+\frac{\sin^2\theta}{\Sigma}
\Bigl[a\,dt-(r^2+a^2)d\phi\Bigr]^2
+\frac{\Sigma}{\Delta}\,dr^2
\\
&-\frac{\Delta}{\Sigma}
\Bigl(dt-a\sin^2\theta\,d\phi\Bigr)^2
\end{aligned}
\label{eq:GDmetric}
\end{equation}
where
\begin{equation}
\begin{aligned}
\Sigma&=r^2+a^2\cos^2\theta,\\
\Delta&=r^2+a^2-2Mr
+(p^2+q^2)\Bigl[1-\beta(r^2+a^2)\Bigr]^2 .
\end{aligned}
\label{eq:SigDelta}
\end{equation}
Here $M$ is the mass, $a$ is the rotation parameter, and $q$ and $p$
are the electric and magnetic charges of the branch under consideration.
Since this is an Einstein--NLED solution, these charges should not be viewed
merely as a formal copy of the Maxwell sector, but as part of the exact
constitutive branch that supports both the electromagnetic field and the
geometry \cite{GarciaDiaz2021,AyonBeato2024}.

In the original presentation of the family, in addition to $\beta$, further
electrodynamic constants usually denoted by $F_0$ and $G_0$ appear and
encode part of the invariant structure of the underlying theory
\cite{GarciaDiaz2021,GarciaDiaz2022Adsds}. In this work these constants are
kept fixed, and we restrict the analysis to the sector
\begin{equation}
(M,a,p,q,\beta).
\label{eq:parameter_sector}
\end{equation}
The parameter $\beta$ controls the nonlinear deformation that appears both
in the electromagnetic potentials and in the radial function $\Delta$.
Since the combination $\beta(r^2+a^2)$ is dimensionless,
\begin{equation}
[\beta]=L^{-2},
\qquad
\ell_\beta\sim |\beta|^{-1/2},
\label{eq:beta_dimension_scale}
\end{equation}
so that $\ell_\beta$ sets the radial scale associated with the
nonlinearity.

Two limits provide immediate checks on the solution. In the linear limit,
\begin{equation}
\Delta\big|_{\beta=0}
=
r^2+a^2-2Mr+(p^2+q^2),
\label{eq:Delta_Maxwell_limit}
\end{equation}
the geometry reduces to the dyonic Kerr--Newman sector
\cite{NewmanEtAl1965,GarciaDiaz2021}. On the other hand, when the rotation
is switched off one obtains
\begin{equation}
ds^2\big|_{a=0}
=
-f(r)\,dt^2+\frac{dr^2}{f(r)}
+r^2(d\theta^2+\sin^2\theta\,d\phi^2),
\label{eq:static_metric_limit}
\end{equation}
with
\begin{equation}
f(r)=
1-\frac{2M}{r}
+
\frac{p^2+q^2}{r^2}\bigl(1-\beta r^2\bigr)^2.
\label{eq:static_lapse_limit}
\end{equation}
The same family therefore contains both the linear dyonic limit and a
nonlinear static sector.

The asymptotic structure requires an additional precaution. From
\eqref{eq:SigDelta}, as $r\to\infty$,
\begin{equation}
\begin{aligned}
\Delta(r)={}&\beta^2(p^2+q^2)r^4
+\bigl[1-2\beta(p^2+q^2)\bigr]r^2
\\
&+2\beta^2a^2(p^2+q^2)r^2
-2Mr+\mathcal O(1).
\end{aligned}
\label{eq:Delta_large_r}
\end{equation}
Therefore, for $Q_c^2:=p^2+q^2>0$ and $\beta\neq0$, this branch is not
asymptotically flat. Indeed, since $\Sigma\sim r^2$, the dominant term
$\Delta\sim \beta^2Q_c^2r^4$ implies
$\Delta/\Sigma\sim \beta^2Q_c^2r^2$ in the far region. This observation
only identifies the leading radial behavior; it is not meant to classify the
solution globally as Kerr--AdS. Extensions of the family may contain
de Sitter- or anti--de Sitter-type sectors, depending on the electrodynamic
branch and on the additional constants
\cite{GarciaDiaz2022Adsds,BretonGutierrezCanoGarciaDiaz2022}. For this
reason, the natural optical description is not based on impact parameters
defined at a flat infinity, but on local directions measured by observers
located at finite distance
\cite{GrenzebachPerlickLammerzahl2014,PerlickTsupko2022}.

\subsection{Mixed potentials and adapted variables}
\label{subsec:potentials_adapted_variables}

The geometry above is supported by an electromagnetic sector which, in the
mixed formulation, is described locally by two potentials
\cite{AyonBeato2024},
\begin{equation}
F=dA,
\qquad
{}^\star P=dA^\star.
\label{eq:F_and_dualP}
\end{equation}
Here $A^\star$ denotes the conjugate potential associated with the
excitation $P_{\mu\nu}$. The symbol $\star$ in $A^\star$ is part of the
notation, whereas in ${}^\star P$ it denotes the Hodge dual of $P$. As
usual in the presence of magnetic charge, these relations must be understood
locally, within a stationary and axisymmetric gauge patch \cite{WuYang1975}.

We write
\begin{equation}
A=A_t\,dt+A_\phi\,d\phi,
\qquad
A^\star=A_t^\star\,dt+A_\phi^\star\,d\phi.
\label{eq:potentials}
\end{equation}
For the García--Díaz branch considered here, the exact potentials are
\cite{GarciaDiaz2021,AyonBeato2024}
\begin{equation}
\begin{aligned}
A={}&
\frac{p\cos\theta\bigl(1-\beta a^2\sin^2\theta\bigr)}{\Sigma}
\Bigl[a\,dt-(r^2+a^2)d\phi\Bigr]
\\
&-
\frac{qr\bigl(1-\beta(r^2+a^2)\bigr)}{\Sigma}
\Bigl(dt-a\sin^2\theta\,d\phi\Bigr) .
\end{aligned}
\label{eq:Afull}
\end{equation}
\begin{equation}
\begin{aligned}
A^\star={}&
\frac{q\cos\theta\bigl(1-\beta a^2\sin^2\theta\bigr)}{\Sigma}
\Bigl[a\,dt-(r^2+a^2)d\phi\Bigr]
\\
&+
\frac{pr\bigl(1-\beta(r^2+a^2)\bigr)}{\Sigma}
\Bigl(dt-a\sin^2\theta\,d\phi\Bigr) .
\end{aligned}
\label{eq:Astarfull}
\end{equation}
In the limit $\beta\to0$, these expressions recover the dyonic potential
compatible with Kerr--Newman. The simultaneous presence of $A$ and
$A^\star$ is not redundant: in the mixed formulation, both contain the
information needed to reconstruct the field--excitation pair $(F,P)$ and,
from it, the local constitutive response \cite{AyonBeato2024}.

The structure of \eqref{eq:Afull} and \eqref{eq:Astarfull} is simplified by
introducing the adapted variables
\begin{equation}
\begin{aligned}
x&:=a\cos\theta,
&
y&:=r,
\\
\delta&:=1-\beta a^2,
&
\Sigma&=x^2+y^2 .
\end{aligned}
\label{eq:xydelta}
\end{equation}
These variables are not new global coordinates. They are local abbreviations
that cleanly separate the angular and radial dependence. In particular, when
$a=0$, the variable $x$ degenerates to $x=0$ and should not be treated
as an independent coordinate.

With these definitions,
\begin{equation}
1-\beta a^2\sin^2\theta=\delta+\beta x^2,
\qquad
1-\beta(r^2+a^2)=\delta-\beta y^2.
\label{eq:kappa_xy_short}
\end{equation}
The temporal components of the potentials then take the compact form
\begin{equation}
\begin{aligned}
A_t&=
\frac{p\,x(\delta+\beta x^2)-q\,y(\delta-\beta y^2)}{\Sigma},\\[1mm]
A_t^\star&=
\frac{q\,x(\delta+\beta x^2)+p\,y(\delta-\beta y^2)}{\Sigma}.
\end{aligned}
\label{eq:AtAtstar}
\end{equation}
In this form, $x=a\cos\theta$ organizes the angular part and $y=r$ the
radial part. Rotation allows the magnetic charge to contribute to the
temporal component and the electric charge to contribute to the azimuthal
component, while the nonlinear deformation is concentrated in the factors
$\delta+\beta x^2$ and $\delta-\beta y^2$.

\subsection{Principal tetrad and aligned fields}
\label{subsec:aligned_scalars}

The line element \eqref{eq:GDmetric} is written in terms of two natural
differential pairs: a temporal--radial pair,
$(dt-a\sin^2\theta\,d\phi,dr)$, and an angular pair,
$(d\theta,(r^2+a^2)d\phi-a\,dt)$. This organization allows one to choose
an orthonormal coframe adapted to the rotating background, analogous to the
one used in the tetrad analysis of Kerr and Kerr--Newman
\cite{Chandrasekhar1983}. For the García--Díaz branch we take
\begin{equation}
\begin{aligned}
\omega^{\hat 0}
&=
\sqrt{\frac{\Delta}{\Sigma}}
\Bigl(dt-a\sin^2\theta\,d\phi\Bigr),\\[1mm]
\omega^{\hat 1}
&=
\sqrt{\frac{\Sigma}{\Delta}}\,dr .
\end{aligned}
\label{eq:tetrad01}
\end{equation}
\begin{equation}
\begin{aligned}
\omega^{\hat 2}
&=
\sqrt{\Sigma}\,d\theta,\\[1mm]
\omega^{\hat 3}
&=
\frac{\sin\theta}{\sqrt{\Sigma}}
\Bigl((r^2+a^2)d\phi-a\,dt\Bigr) .
\end{aligned}
\label{eq:tetrad23}
\end{equation}
In the region
\begin{equation}
\Delta(r)>0,
\qquad
\Sigma>0,
\qquad
0<\theta<\pi,
\label{eq:tetrad_regular_domain}
\end{equation}
this coframe satisfies
\begin{equation}
ds^2=
-(\omega^{\hat0})^2
+(\omega^{\hat1})^2
+(\omega^{\hat2})^2
+(\omega^{\hat3})^2.
\label{eq:metric_tetrad}
\end{equation}
We also fix the orientation of the orthonormal frame by
\begin{equation}
\omega^{\hat0}\wedge\omega^{\hat1}\wedge
\omega^{\hat2}\wedge\omega^{\hat3}>0.
\label{eq:orientation}
\end{equation}
With the signature \eqref{eq:signature_convention}, this choice determines
the action of the Hodge dual on the two principal planes,
\begin{equation}
{}^\star(\omega^{\hat0}\wedge\omega^{\hat1})
=
-\omega^{\hat2}\wedge\omega^{\hat3},
\qquad
{}^\star(\omega^{\hat2}\wedge\omega^{\hat3})
=
\omega^{\hat0}\wedge\omega^{\hat1}.
\label{eq:Hodge_basis_actions}
\end{equation}
These conventions fix the relative signs that will appear when extracting
the excitation $P$ from the conjugate potential $A^\star$.

The usefulness of this frame is not only metric. In the mixed formulation of
Plebański and Salazar--García--Plebański, the field and the excitation take
an especially simple form when projected onto the principal tetrad
\cite{Plebanski1970,SalazarGarciaPlebanski1987,AyonBeato2024}. In this
frame they are aligned as
\begin{equation}
\begin{aligned}
F&=
E\,\omega^{\hat0}\wedge\omega^{\hat1}
+B\,\omega^{\hat2}\wedge\omega^{\hat3},\\
P&=
D\,\omega^{\hat0}\wedge\omega^{\hat1}
+H\,\omega^{\hat2}\wedge\omega^{\hat3} .
\end{aligned}
\label{eq:FPaligned}
\end{equation}
The scalars $E$ and $B$ are the electric and magnetic components of the
field $F$ in the principal frame, while $D$ and $H$ are the
corresponding components of the excitation $P$.

The adapted variables allow these four scalars to be extracted directly from
the temporal components of the potentials. Indeed,
\begin{equation}
\begin{aligned}
dx&=-a\sin\theta\,d\theta,
&
 dy&=dr,\\
\partial_\theta&=-a\sin\theta\,\partial_x,
&
\partial_r&=\partial_y .
\end{aligned}
\label{eq:xy_derivatives}
\end{equation}
Furthermore,
\begin{equation}
\omega^{\hat0}\wedge\omega^{\hat1}
=
\Bigl(dt-a\sin^2\theta\,d\phi\Bigr)\wedge dr,
\label{eq:wedge01_explicit}
\end{equation}
whereas
\begin{equation}
\omega^{\hat2}\wedge\omega^{\hat3}
=
\sin\theta\,d\theta\wedge
\Bigl[(r^2+a^2)d\phi-a\,dt\Bigr].
\label{eq:wedge23_explicit}
\end{equation}
Thus, comparing the coefficients of $dt\wedge dy$ and $dx\wedge dt$ in
$dA$ fixes the components of the field $F$. For the excitation, by
contrast, one compares $dA^\star$ with ${}^\star P$. From
\eqref{eq:Hodge_basis_actions}, if
$P=D\,\omega^{\hat0}\wedge\omega^{\hat1}
+H\,\omega^{\hat2}\wedge\omega^{\hat3}$, then
\[
{}^\star P
=
-D\,\omega^{\hat2}\wedge\omega^{\hat3}
+
H\,\omega^{\hat0}\wedge\omega^{\hat1}.
\]
With these conventions one obtains
\begin{equation}
\begin{aligned}
E&=-\partial_y A_t,
&
B&=\partial_x A_t,\\
D&=-\partial_x A_t^\star,
&
H&=-\partial_y A_t^\star .
\end{aligned}
\label{eq:EDBHdef}
\end{equation}
These relations will be used to translate the exact potentials into the
aligned scalars of the background.

Substituting \eqref{eq:AtAtstar} into \eqref{eq:EDBHdef}, one obtains
\begin{equation}
\begin{aligned}
B(x,y)
={}&
\frac{1}{\Sigma^2}\Bigl\{
\delta\Bigl[p(y^2-x^2)+2qxy\Bigr]
\\
&+\beta\Bigl[p\,x^2(x^2+3y^2)-2qxy^3\Bigr]\Bigr\} .
\end{aligned}
\label{eq:Bxy}
\end{equation}
\begin{equation}
\begin{aligned}
E(x,y)
={}&
\frac{1}{\Sigma^2}\Bigl\{
\delta\Bigl[q(x^2-y^2)+2pxy\Bigr]
\\
&+\beta\Bigl[2px^3y-qy^2(3x^2+y^2)\Bigr]\Bigr\} .
\end{aligned}
\label{eq:Exy}
\end{equation}
\begin{equation}
\begin{aligned}
D(x,y)
={}&
\frac{1}{\Sigma^2}\Bigl\{
\delta\Bigl[q(x^2-y^2)+2pxy\Bigr]
\\
&-\beta\Bigl[2pxy^3+q\,x^2(x^2+3y^2)\Bigr]\Bigr\} .
\end{aligned}
\label{eq:Dxy}
\end{equation}
\begin{equation}
\begin{aligned}
H(x,y)
={}&
\frac{1}{\Sigma^2}\Bigl\{
\delta\Bigl[p(y^2-x^2)+2qxy\Bigr]
\\
&+\beta\Bigl[p\,y^2(3x^2+y^2)+2qx^3y\Bigr]\Bigr\} .
\end{aligned}
\label{eq:Hxy}
\end{equation}
These expressions are exact. In the linear limit $\beta\to0$, one recovers
$D=E$ and $H=B$. Thus the terms proportional to $\beta$ are precisely
the ones that separate field and excitation and anticipate a nontrivial
constitutive response.

\subsection{Algebraic identities for optics}
\label{subsec:optical_algebra}

With the tetrad conventions fixed in
\eqref{eq:orientation}--\eqref{eq:Hodge_basis_actions}, the signs of the
pseudoscalar invariants are determined. For the field $F_{\mu\nu}$ we use
the standard NLED invariants
\begin{equation}
\mathcal F:=\frac14 F_{\mu\nu}F^{\mu\nu},
\qquad
\mathcal G:=\frac14F_{\mu\nu}\,{}^\star F^{\mu\nu}.
\label{eq:FGdef}
\end{equation}
The analysis of optical characteristics in NLED is formulated precisely in
terms of these invariants and of the constitutive response
\cite{Boillat1970,ObukhovRubilar2002}. Using the aligned form
\eqref{eq:FPaligned}, one obtains
\begin{equation}
\mathcal F
=
-\frac12(E^2-B^2),
\qquad
\mathcal G=-EB.
\label{eq:FGEB}
\end{equation}
For the excitation we define analogously
\begin{equation}
\mathcal P:=\frac14 P_{\mu\nu}P^{\mu\nu},
\qquad
\mathcal Q:=\frac14P_{\mu\nu}\,{}^\star P^{\mu\nu}.
\label{eq:PQdef}
\end{equation}
so that
\begin{equation}
\mathcal P
=
-\frac12(D^2-H^2),
\qquad
\mathcal Q=-DH.
\label{eq:PQDH}
\end{equation}
The same information can be condensed in the complex form
\begin{equation}
\mathcal F+i\mathcal G
=
-\frac12(E+iB)^2,
\qquad
\mathcal P+i\mathcal Q
=
-\frac12(D+iH)^2.
\label{eq:complexinvariants}
\end{equation}
These identities summarize the structure of the invariants in the principal
frame and fix the conventions that will be kept throughout the optical
construction.

In addition to the scalar invariants, the effective optics will depend on the
quadratic tensor selected by the background field,
\begin{equation}
t^{\mu\nu}:=F^\mu{}_{\alpha}F^{\alpha\nu}.
\label{eq:tmunu_background_def}
\end{equation}
In the orthonormal frame, the aligned form implies that the only nonzero
blocks of $F_{\hat a\hat b}$ lie in the $(\hat0,\hat1)$ and
$(\hat2,\hat3)$ planes. Raising indices with
$\eta_{\hat a\hat b}=\mathrm{diag}(-1,1,1,1)$, one obtains
\begin{equation}
t^{\hat a\hat b}
=
\mathrm{diag}
\bigl(
-E^2,\,
+E^2,\,
-B^2,\,
-B^2
\bigr).
\label{eq:tmunu_tetrad_correct}
\end{equation}
The electric field distinguishes the temporal--radial plane, while the
magnetic field distinguishes the angular plane. This block separation is the
algebraic imprint left by the electromagnetic background on optical
propagation in NLED \cite{NovelloEtAl2000,ObukhovRubilar2002}.

The difference between field and excitation is read directly from
\eqref{eq:Bxy}--\eqref{eq:Hxy}. Subtracting the corresponding components,
the terms proportional to $\delta$ cancel and one obtains
\begin{equation}
\begin{aligned}
D-E
&=
-\beta\frac{q(x^2-y^2)+2pxy}{\Sigma},
\\
H-B
&=
\beta\frac{p(y^2-x^2)+2qxy}{\Sigma} .
\end{aligned}
\label{eq:DEHB}
\end{equation}
Therefore
\begin{equation}
D\to E,
\qquad
H\to B,
\qquad
(\beta\to0),
\label{eq:Maxwelllimitfields}
\end{equation}
as required by the linear limit.

The same combinations that appeared in the potentials are naturally grouped
into two factors, one radial and one angular,
\begin{equation}
\kappa_r(r):=1-\beta(r^2+a^2),
\qquad
\kappa_\theta(\theta):=1-\beta a^2\sin^2\theta.
\label{eq:kappas}
\end{equation}
The regular branch connected with Maxwell is characterized by
\begin{equation}
\kappa_r(r)\neq0,
\qquad
\kappa_\theta(\theta)>0.
\label{eq:kappa_regular_domain}
\end{equation}
These conditions do not impose an additional dynamical restriction on the
geometry; they only delimit the local chart in which the pair $(D,B)$ can
be used as constitutive variables. Within this chart, the Maxwell limit is
reached without singularities, and the local relation between field and
excitation is prepared for the reconstruction carried out in the next
section.


\section{Local constitutive branch and response matrix}
\label{sec:constitutive}

The aligned fields found above reduce the electromagnetic sector to a local
constitutive question.  We have to understand how the excitation pair
$(D,B)$ is related to the field pair $(E,H)$.  In Maxwell theory this relation
is simply
\begin{equation}
D=E,
\qquad
H=B,
\label{eq:Maxwell_constitutive_limit_III}
\end{equation}
but Eq.~\eqref{eq:DEHB} shows that this equality is deformed when
$\beta\neq0$.  Thus the local optical response of the exact solution is
encoded in the map
\begin{equation}
(D,B)\longmapsto(E,H),
\label{eq:constitutive_map_intro}
\end{equation}
inside the regular branch connected with Maxwell theory.

The mixed formulation of Pleba\'nski and Salazar--Garc\'ia--Pleba\'nski
allows this map to be described by local structural functions, without
assuming a single global elementary Lagrangian $L(\mathcal F,\mathcal G)$
\cite{Plebanski1970,SalazarGarciaPlebanski1987,SalazarGarciaPlebanski1989}.
For the rotating Garc\'ia--D\'iaz solution, this is precisely the formulation
that identifies the underlying electrodynamics \cite{AyonBeato2024}.  In the
following, no external material law is introduced.  We only reorganize the
exact background fields in the local chart that they naturally define.

We work in a connected domain defined by
\begin{equation}
Q_c:=\sqrt{p^2+q^2}>0,
\qquad
\kappa_r(r)\neq0,
\qquad
\kappa_\theta(\theta)>0.
\label{eq:III_regular_domain}
\end{equation}
The first condition removes the electromagnetically trivial case.  The second
keeps the $(D,B)$ chart regular, and the third selects the angular subbranch
that is smoothly connected with Maxwell theory.  In each connected component
with $\kappa_r\neq0$, we write
\begin{equation}
\varepsilon_r:=\operatorname{sgn}\kappa_r=\pm1.
\label{eq:epsr_def}
\end{equation}
This sign is fixed inside the chart.

\subsection{Mixed structural functions and exact background relation}
\label{subsec:mixed_structural_functions}

In the mixed formulation, the constitutive response can be encoded locally by
two structural functions
\cite{Plebanski1970,SalazarGarciaPlebanski1987},
\begin{equation}
\mathcal M^{(+)}=\mathcal M^{(+)}(D,B),
\qquad
\mathcal M^{(-)}=\mathcal M^{(-)}(E,H),
\label{eq:Mpm_def}
\end{equation}
whose differentials satisfy
\begin{equation}
d\mathcal M^{(+)}
=
E\,dD+H\,dB,
\qquad
d\mathcal M^{(-)}
=
D\,dE+B\,dH.
\label{eq:dMpm}
\end{equation}
Thus
\begin{equation}
E=\partial_D\mathcal M^{(+)},
\qquad
H=\partial_B\mathcal M^{(+)},
\label{eq:Mplus_derivatives}
\end{equation}
and
\begin{equation}
D=\partial_E\mathcal M^{(-)},
\qquad
B=\partial_H\mathcal M^{(-)}.
\label{eq:Mminus_derivatives}
\end{equation}
The signs $+$ and $-$ in $\mathcal M^{(\pm)}$ label conjugate
constitutive representations. They should not be confused with the two
optical branches that will appear when the Fresnel quartic is factorized.

To condense the exact fields obtained in the previous section, we introduce
the combinations
\begin{equation}
\mathcal A:=q(x^2-y^2)+2pxy,
\qquad
\mathcal C:=p(y^2-x^2)+2qxy.
\label{eq:AC_def}
\end{equation}
These quantities are the structures that survive in the linear limit and
organize, respectively, the electric and magnetic sectors of the aligned
field. In terms of them, the four scalars $D,B,E,H$ admit a simple
factorization after constant charge-dependent shifts. These shifts are not
arbitrary: they are fixed by the requirement that the pair $(D,B)$ share a
single radial factor, whereas the pair $(E,H)$ share a single angular
factor.

The essential step can be seen in the component $D$. From
\eqref{eq:Dxy},
\begin{equation}
D
=
\frac{
\delta\mathcal A
-
\beta\left[
2pxy^3+q\,x^2(x^2+3y^2)
\right]
}{\Sigma^2},
\label{eq:D_exact_rewritten_III}
\end{equation}
one has
\begin{equation}
D+\beta q
=
\frac{
\delta\mathcal A
-
\beta\left[
2pxy^3+q\,x^2(x^2+3y^2)
\right]
+\beta q\Sigma^2
}{\Sigma^2}.
\label{eq:D_shift_intermediate_III}
\end{equation}
The added term combines with the terms proportional to $\beta$. Using
$\Sigma^2=(x^2+y^2)^2$, the numerator reduces to
\begin{equation}
\delta\mathcal A-\beta y^2\mathcal A
=
(\delta-\beta y^2)\mathcal A
=
\kappa_r\mathcal A.
\label{eq:D_shift_factorization_III}
\end{equation}
Therefore,
\begin{equation}
D+\beta q
=
\frac{\kappa_r\mathcal A}{\Sigma^2}.
\label{eq:D_shifted_exact_III}
\end{equation}
The same computation applied to the magnetic component gives
\begin{equation}
B-\beta p
=
\frac{\kappa_r\mathcal C}{\Sigma^2}.
\label{eq:B_shifted_exact_III}
\end{equation}
Thus the pair $(D,B)$ is controlled by a single radial factor:
\begin{equation}
D+\beta q
=
\frac{\kappa_r\mathcal A}{\Sigma^2},
\qquad
B-\beta p
=
\frac{\kappa_r\mathcal C}{\Sigma^2}.
\label{eq:DB_shifted_exact}
\end{equation}

Analogously, the pair $(E,H)$ is controlled by the angular factor:
\begin{equation}
E+\beta q
=
\frac{\kappa_\theta\mathcal A}{\Sigma^2},
\qquad
H-\beta p
=
\frac{\kappa_\theta\mathcal C}{\Sigma^2}.
\label{eq:EH_shifted_exact}
\end{equation}
These relations motivate the introduction of the shifted vectors
\begin{equation}
\mathbf U_+
:=
\begin{pmatrix}
D+\beta q\\
B-\beta p
\end{pmatrix},
\qquad
\mathbf U_-
:=
\begin{pmatrix}
E+\beta q\\
H-\beta p
\end{pmatrix}.
\label{eq:Uplus_def}
\end{equation}
The sign $+$ refers to the direct representation in the $(D,B)$ plane,
whereas the sign $-$ refers to the conjugate representation in the
$(E,H)$ plane. This notation is purely constitutive.

We also define the associated norms
\begin{equation}
\begin{aligned}
\rho_+&:=\|\mathbf U_+\|
=\sqrt{(D+\beta q)^2+(B-\beta p)^2},\\[1mm]
\rho_-&:=\|\mathbf U_-\|
=\sqrt{(E+\beta q)^2+(H-\beta p)^2} .
\end{aligned}
\label{eq:rho_pm_def_after_U}
\end{equation}
To evaluate them on the background, we use the identity
\begin{equation}
\mathcal A^2+\mathcal C^2=Q_c^2\Sigma^2.
\label{eq:AC_norm_identity}
\end{equation}
A direct way to see this is to write
\begin{equation}
X:=x^2-y^2,
\qquad
Y:=2xy.
\label{eq:XY_def_III}
\end{equation}
Then
\begin{equation}
\mathcal A=qX+pY,
\qquad
\mathcal C=-pX+qY,
\label{eq:AC_XY_form_III}
\end{equation}
and hence
\begin{equation}
\mathcal A^2+\mathcal C^2
=
(p^2+q^2)(X^2+Y^2)
=
Q_c^2(x^2+y^2)^2
=
Q_c^2\Sigma^2.
\label{eq:AC_norm_derivation_III}
\end{equation}
From \eqref{eq:DB_shifted_exact} and \eqref{eq:EH_shifted_exact}, it follows
that
\begin{equation}
\begin{aligned}
\rho_+^2
&=
\frac{Q_c^2\kappa_r^2}{\Sigma^2},\\[1mm]
\rho_-^2
&=
\frac{Q_c^2\kappa_\theta^2}{\Sigma^2}.
\end{aligned}
\label{eq:rho_pm_closed}
\end{equation}

Eliminating $\mathcal A$ and $\mathcal C$ between
\eqref{eq:DB_shifted_exact} and \eqref{eq:EH_shifted_exact}, the background
satisfies
\begin{equation}
E+\beta q
=
\frac{\kappa_\theta}{\kappa_r}(D+\beta q),
\qquad
H-\beta p
=
\frac{\kappa_\theta}{\kappa_r}(B-\beta p).
\label{eq:constitutive_shifted_exact}
\end{equation}
Since
\begin{equation}
\kappa_\theta-\kappa_r=\beta\Sigma,
\label{eq:kappa_difference}
\end{equation}
the same relation can be written as
\begin{equation}
E
=
D+\frac{\beta\Sigma}{\kappa_r}(D+\beta q),
\qquad
H
=
B+\frac{\beta\Sigma}{\kappa_r}(B-\beta p).
\label{eq:constitutive_exact_kappa}
\end{equation}
This equation does not postulate an independent law: it is the same
electromagnetic information of the background, reorganized as a local
constitutive relation between excitation and field.

\subsection{Local constitutive continuation}
\label{subsec:local_continuation}

To turn \eqref{eq:constitutive_exact_kappa} into a local chart written only
in terms of $(D,B)$, the geometric factor $\Sigma/\kappa_r$ must be
eliminated in favor of a quantity defined on the constitutive plane. The norm
of $\mathbf U_+$ provides precisely this quantity. On the background,
\eqref{eq:rho_pm_closed} implies
\begin{equation}
\rho_+
=
\frac{Q_c|\kappa_r|}{\Sigma}.
\label{eq:rho_plus_background}
\end{equation}
Since we work in a connected component where $\kappa_r\neq0$, the sign
$\varepsilon_r=\operatorname{sgn}\kappa_r$ is constant. Therefore,
\begin{equation}
|\kappa_r|=\varepsilon_r\kappa_r,
\qquad
\varepsilon_r^2=1,
\label{eq:abs_kappar_epsr}
\end{equation}
and
\begin{equation}
\rho_+
=
\frac{Q_c\,\varepsilon_r\kappa_r}{\Sigma}.
\label{eq:rho_plus_eps_form}
\end{equation}
It follows that
\begin{equation}
\frac{\Sigma}{\kappa_r}
=
\varepsilon_r\,\frac{Q_c}{\rho_+},
\qquad
\frac{\beta\Sigma}{\kappa_r}
=
\varepsilon_r\,\frac{\beta Q_c}{\rho_+}.
\label{eq:betaSigma_over_kappar}
\end{equation}
Substituting this identity into \eqref{eq:constitutive_exact_kappa} gives the
local representative of the direct branch:
\begin{equation}
E
=
D+\varepsilon_r\,\beta Q_c\,\frac{D+\beta q}{\rho_+},
\qquad
H
=
B+\varepsilon_r\,\beta Q_c\,\frac{B-\beta p}{\rho_+}.
\label{eq:EH_constitutive_explicit}
\end{equation}
Here $\rho_+=\|\mathbf U_+\|$ is understood as a function of the
constitutive variables $(D,B)$, not only as a background quantity. The
local chart requires
\begin{equation}
\rho_+\neq0,
\label{eq:rho_plus_nonzero_local}
\end{equation}
because the shifted origin $(D+\beta q,B-\beta p)=(0,0)$ makes the
direction of $\mathbf U_+$ singular. On the background, and for $Q_c>0$,
this condition reduces exactly to $\kappa_r\neq0$.

The continuation \eqref{eq:EH_constitutive_explicit} is integrable in the
$(D,B)$ plane and therefore admits a local structural function. Up to an
irrelevant additive constant,
\begin{equation}
\mathcal M^{(+)}(D,B)
=
\frac12(D^2+B^2)
+
\varepsilon_r\beta Q_c\,\rho_+.
\label{eq:Mplus_local_explicit}
\end{equation}
Indeed,
\begin{equation}
\partial_D\rho_+
=
\frac{D+\beta q}{\rho_+},
\qquad
\partial_B\rho_+
=
\frac{B-\beta p}{\rho_+},
\label{eq:rho_plus_derivatives}
\end{equation}
and therefore
\begin{equation}
\begin{aligned}
\partial_D\mathcal M^{(+)}
&=
D+\varepsilon_r\beta Q_c\frac{D+\beta q}{\rho_+},
\\
\partial_B\mathcal M^{(+)}
&=
B+\varepsilon_r\beta Q_c\frac{B-\beta p}{\rho_+} .
\end{aligned}
\label{eq:Mplus_derivatives_local_check}
\end{equation}
Thus,
\begin{equation}
E=\partial_D\mathcal M^{(+)},
\qquad
H=\partial_B\mathcal M^{(+)}.
\label{eq:Mplus_local_reproduces_EH}
\end{equation}
The response matrix
\[
\mathbb C
=
\frac{\partial(E,H)}{\partial(D,B)}
\]
is then the local Hessian of $\mathcal M^{(+)}$. In particular, the
equality of mixed derivatives,
$\partial_B E=\partial_D H$, expresses the integrability of this
constitutive representation.

In the linear limit,
\begin{equation}
\beta\to0,
\qquad
E\to D,
\qquad
H\to B,
\label{eq:EH_Maxwell_limit}
\end{equation}
the structural function reduces to the Maxwell quadratic form,
$\mathcal M^{(+)}\to\tfrac12(D^2+B^2)$, and the constitutive anisotropy
disappears.

\subsection{Response matrix and spectral decomposition}
\label{subsec:constitutive_matrix}

The differential response of the local branch is obtained by linearizing the
chart \eqref{eq:EH_constitutive_explicit} with $p,q,\beta$, and
$\varepsilon_r$ fixed. We define
\begin{equation}
\mathbb C
:=
\frac{\partial(E,H)}{\partial(D,B)}
=
\begin{pmatrix}
\partial_D E & \partial_B E\\[1mm]
\partial_D H & \partial_B H
\end{pmatrix}.
\label{eq:C_def}
\end{equation}
By the construction above, $\mathbb C$ coincides with the local Hessian of
$\mathcal M^{(+)}$.

All the nonlinear dependence of \eqref{eq:EH_constitutive_explicit} enters
through the normalized direction $\mathbf U_+/\rho_+$. Since
\begin{equation}
\begin{aligned}
\rho_+&=(\mathbf U_+^T\mathbf U_+)^{1/2},\\
d\rho_+&=
\frac{\mathbf U_+^T d\mathbf U_+}{\rho_+} .
\end{aligned}
\label{eq:rho_plus_differential}
\end{equation}
one obtains
\begin{equation}
d\!\left(\frac{\mathbf U_+}{\rho_+}\right)
=
\left(
\frac{\mathbb I}{\rho_+}
-
\frac{\mathbf U_+\mathbf U_+^T}{\rho_+^3}
\right)d\mathbf U_+,
\label{eq:normalized_vector_identity}
\end{equation}
where $\mathbb I$ is the two-dimensional identity. Since $\mathbf U_+$
differs from $(D,B)^T$ only by a constant shift,
$d\mathbf U_+=(dD,dB)^T$. In this way,
\begin{equation}
\mathbb C
=
\mathbb I
+
\varepsilon_r\,\beta Q_c
\left(
\frac{\mathbb I}{\rho_+}
-
\frac{\mathbf U_+\mathbf U_+^T}{\rho_+^3}
\right).
\label{eq:C_compact}
\end{equation}
In components,
\begin{widetext}
\begin{equation}
\mathbb C
=
\begin{pmatrix}
1+\dfrac{\varepsilon_r\beta Q_c(B-\beta p)^2}{\rho_+^3}
&
-\dfrac{\varepsilon_r\beta Q_c(D+\beta q)(B-\beta p)}{\rho_+^3}
\\[3mm]
-\dfrac{\varepsilon_r\beta Q_c(D+\beta q)(B-\beta p)}{\rho_+^3}
&
1+\dfrac{\varepsilon_r\beta Q_c(D+\beta q)^2}{\rho_+^3}
\end{pmatrix} .
\label{eq:C_components}
\end{equation}
\end{widetext}

The compact form \eqref{eq:C_compact} shows that the response distinguishes a
special direction in the constitutive plane. We define
\begin{equation}
\hat{\mathbf n}_+
:=
\frac{\mathbf U_+}{\rho_+},
\qquad
\|\hat{\mathbf n}_+\|=1,
\label{eq:nplus_unit}
\end{equation}
and the projectors
\begin{equation}
\Pi_\parallel
:=
\hat{\mathbf n}_+\hat{\mathbf n}_+^T,
\qquad
\Pi_\perp
:=
\mathbb I-\Pi_\parallel.
\label{eq:projectors_def}
\end{equation}
Any variation $\delta\mathbf u=(\delta D,\delta B)^T$ decomposes as
\begin{equation}
\delta\mathbf u
=
\delta\mathbf u_\parallel+\delta\mathbf u_\perp,
\qquad
\delta\mathbf u_\parallel:=\Pi_\parallel\delta\mathbf u,
\qquad
\delta\mathbf u_\perp:=\Pi_\perp\delta\mathbf u.
\label{eq:parallel_perp_decomposition}
\end{equation}
Here ``parallel'' and ``perpendicular'' refer only to directions within the
$(D,B)$ plane; they are not yet optical polarizations.

In terms of these projectors,
\begin{equation}
\frac{\mathbb I}{\rho_+}
-
\frac{\mathbf U_+\mathbf U_+^T}{\rho_+^3}
=
\frac{1}{\rho_+}
\left(
\mathbb I-
\frac{\mathbf U_+\mathbf U_+^T}{\rho_+^2}
\right)
=
\frac{1}{\rho_+}\Pi_\perp.
\label{eq:C_projector_identity}
\end{equation}
Since $\mathbb I=\Pi_\parallel+\Pi_\perp$, the response matrix takes the
spectral form
\begin{equation}
\mathbb C
=
\Pi_\parallel+\chi\,\Pi_\perp,
\qquad
\chi
:=
1+\varepsilon_r\,\frac{\beta Q_c}{\rho_+}.
\label{eq:C_spectral}
\end{equation}
The eigenvalues are
\begin{equation}
\lambda_\parallel=1,
\qquad
\lambda_\perp=\chi.
\label{eq:eigenvalues_C}
\end{equation}
Thus, a perturbation parallel to $\mathbf U_+$ remains undeformed by the
constitutive response, whereas the perpendicular component is multiplied by
$\chi$. If
\begin{equation}
\delta\mathbf v
=
\begin{pmatrix}
\delta E\\
\delta H
\end{pmatrix},
\label{eq:v_variation}
\end{equation}
then the linearized law is
\begin{equation}
\delta\mathbf v
=
\mathbb C\,\delta\mathbf u
=
\delta\mathbf u_\parallel
+
\chi\,\delta\mathbf u_\perp.
\label{eq:linearized_law_split}
\end{equation}
The decomposition above is, at this stage, purely constitutive: it separates
directions in the $(D,B)$ plane, not optical polarizations. The latter are
defined only after inserting $\mathbb C$ into the Fresnel characteristic
problem.

\subsection{Evaluation on the background and domain of validity}
\label{subsec:chi_domain}

Evaluating $\chi$ on the exact background, we use
\eqref{eq:rho_plus_background} together with
\eqref{eq:betaSigma_over_kappar}. Then
\begin{equation}
\chi(r,\theta)
=
1+\frac{\beta\Sigma}{\kappa_r}.
\label{eq:chi_intermediate}
\end{equation}
Using \eqref{eq:kappa_difference}, one obtains
\begin{equation}
\chi(r,\theta)
=
\frac{\kappa_\theta}{\kappa_r}
=
\frac{1-\beta a^2\sin^2\theta}
{1-\beta(r^2+a^2)}.
\label{eq:chi_GD}
\end{equation}
This equality is exact on the background; it is not a perturbative
expansion. In the Maxwell limit,
\begin{equation}
\chi\to1,
\qquad
\lambda_\parallel,\lambda_\perp\to1,
\qquad
(\beta\to0),
\label{eq:chi_Maxwell_limit}
\end{equation}
the response becomes isotropic again in the constitutive plane.

The domain of the chart is fixed by two requirements. First, the
parametrization in terms of $(D,B)$ requires
\begin{equation}
\rho_+\neq0,
\label{eq:rho_nonzero_condition}
\end{equation}
which, on the background and for $Q_c>0$, is equivalent to
\begin{equation}
\kappa_r(r)\neq0.
\label{eq:kappa_r_nonzero}
\end{equation}
Second, the differential response must be invertible. From
\eqref{eq:C_spectral},
\begin{equation}
\det\mathbb C=\chi,
\label{eq:detC_chi}
\end{equation}
and therefore, on the background,
\begin{equation}
\det\mathbb C
=
\frac{\kappa_\theta}{\kappa_r}.
\label{eq:detC_background}
\end{equation}
Within a branch with $\kappa_r\neq0$, invertibility fails if
$\kappa_\theta=0$. The subbranch connected with Maxwell is selected by
\begin{equation}
\kappa_\theta(\theta)>0,
\label{eq:kappatheta_positive_final_III}
\end{equation}
since
\begin{equation}
\kappa_\theta(\theta)
=
1-\beta a^2\sin^2\theta
\longrightarrow 1
\qquad
(\beta\to0).
\label{eq:kappatheta_Maxwell_connection}
\end{equation}

The local constitutive response of the rotating García--Díaz background is
therefore defined in the connected region
\begin{equation}
Q_c>0,
\qquad
\kappa_r(r)\neq0,
\qquad
\kappa_\theta(\theta)>0.
\label{eq:constitutive_domain_summary}
\end{equation}
In that domain,
\begin{equation}
\mathbb C
=
\Pi_\parallel+\chi\,\Pi_\perp,
\qquad
\chi=\frac{\kappa_\theta}{\kappa_r}.
\label{eq:constitutive_summary}
\end{equation}
The background induces an undeformed parallel sector and a perpendicular
sector multiplied by $\kappa_\theta/\kappa_r$. This anisotropy does not
come from an additional approximation: it is the differential form of the
constitutive chart that reproduces the exact background fields. The matrix
$\mathbb C$ will be the local datum entering the Fresnel analysis, where it
is determined whether the constitutive anisotropy becomes a splitting of
optical cones.


\section{Fresnel cones and effective optical metrics}
\label{sec:opticalmetrics}

The response matrix of the previous section is still a local constitutive
object.  It tells us how $(E,H)$ changes when $(D,B)$ is varied, but it does
not yet give the propagation cones.  To obtain the optical geometry, this
response must be inserted into the high-frequency characteristic problem.  In
NLED, the characteristic equation is generally a Fresnel quartic.  When this
quartic factorizes, each quadratic factor defines one effective optical cone
\cite{Boillat1970,NovelloEtAl2000,ObukhovRubilar2002}.

For the rotating Garc\'ia--D\'iaz branch this step is local.  The exact
electrodynamics is most naturally described in mixed variables, and the
supporting Lagrangian need not be available as a single elementary function of
the standard invariants \cite{AyonBeato2024}.  Therefore, the Lagrangian used
below is not a new theory imposed on the solution.  It is a local
representative whose only task is to reproduce, at the perturbative order
considered here, the constitutive derivatives needed for the Fresnel
construction.

\subsection{Local Lagrangian representative}
\label{subsec:local_lagrangian_branch}

We work in the regular constitutive domain defined by
\eqref{eq:constitutive_domain_summary}. To pass to a local Lagrangian
description, it is convenient to use the conjugate representation, in which
$(E,H)$ are taken as independent variables and $(D,B)$ as the response.
We therefore recall the shifted vector of the inverse chart,
\begin{equation}
\begin{aligned}
\mathbf U_-
&=
\begin{pmatrix}
E+\beta q\\
H-\beta p
\end{pmatrix},
\\
\rho_-&=\|\mathbf U_-\|
=\sqrt{(E+\beta q)^2+(H-\beta p)^2} .
\end{aligned}
\label{eq:Uminus_rhominus_recall_IV}
\end{equation}
On the exact background, \eqref{eq:rho_pm_closed} gives
\begin{equation}
\rho_-=
\frac{Q_c\kappa_\theta}{\Sigma}>0,
\label{eq:rho_minus_background_IV}
\end{equation}
where we have used the fact that the regular branch satisfies
$\kappa_\theta>0$. This choice selects the subbranch smoothly connected
with Maxwell theory, since $\kappa_\theta\to1$ as $\beta\to0$.

This is why the inverse representation does not contain the sign
$\varepsilon_r$ that appeared in the direct chart. Indeed, in
\eqref{eq:betaSigma_over_kappar} it was necessary to write
$|\kappa_r|=\varepsilon_r\kappa_r$, because the norm $\rho_+$ contains
$|\kappa_r|$. By contrast, the conjugate norm satisfies
$\rho_-=Q_c\kappa_\theta/\Sigma$ inside the subbranch
$\kappa_\theta>0$. Therefore,
\begin{equation}
\frac{\Sigma}{\kappa_\theta}
=
\frac{Q_c}{\rho_-},
\label{eq:sigma_over_kappatheta_IV}
\end{equation}
without introducing any additional sign. Using also
\begin{equation}
\kappa_r=\kappa_\theta-\beta\Sigma,
\label{eq:kappar_from_kappatheta_IV}
\end{equation}
one obtains
\begin{equation}
\frac{\kappa_r}{\kappa_\theta}
=
1-\frac{\beta\Sigma}{\kappa_\theta}
=
1-\frac{\beta Q_c}{\rho_-}.
\label{eq:kappas_inverse_ratio_IV}
\end{equation}
Thus, by inverting the exact shifted relation
\eqref{eq:constitutive_shifted_exact}, one finds
\begin{equation}
D
=
E-\beta Q_c\,\frac{E+\beta q}{\rho_-},
\qquad
B
=
H-\beta Q_c\,\frac{H-\beta p}{\rho_-}.
\label{eq:DB_inverse_local_exact_IV}
\end{equation}

The mixed structural function $\mathcal M^{(-)}(E,H)$ encodes the same
constitutive information as $\mathcal M^{(+)}(D,B)$, but in the inverse
representation. It should not yet be confused with the spacetime Lagrangian
$L(\mathcal F,\mathcal G)$. Its role is to provide the local map from
which we will reconstruct the derivatives $L_{\mathcal F}$ and
$L_{\mathcal G}$. By definition,
\begin{equation}
d\mathcal M^{(-)}
=
D\,dE+B\,dH .
\label{eq:dMminus_IV}
\end{equation}
Using \eqref{eq:DB_inverse_local_exact_IV} and retaining terms up to first
order in $\beta$, one has
\begin{equation}
\begin{aligned}
D
&=
E-\beta Q_c\frac{E+\beta q}{\rho_-}
+\mathcal O(\beta^2),
\\
B
&=
H-\beta Q_c\frac{H-\beta p}{\rho_-}
+\mathcal O(\beta^2) .
\end{aligned}
\label{eq:DB_from_inverse_chart_IV}
\end{equation}
Moreover,
\begin{equation}
d\rho_-
=
\frac{E+\beta q}{\rho_-}\,dE
+
\frac{H-\beta p}{\rho_-}\,dH .
\label{eq:drhominus_IV}
\end{equation}
Therefore,
\begin{equation}
D\,dE+B\,dH
=
E\,dE+H\,dH
-\beta Q_c\,d\rho_-
+\mathcal O(\beta^2).
\label{eq:dMminus_integrand_IV}
\end{equation}
After integration, and omitting an irrelevant additive constant, one obtains
\begin{equation}
\mathcal M^{(-)}(E,H)
=
\frac12(E^2+H^2)
-\beta Q_c\,\rho_-
+\mathcal O(\beta^2).
\label{eq:Mminus_reduced_IV}
\end{equation}
Its derivatives reproduce, by construction,
\begin{equation}
D=\partial_E\mathcal M^{(-)},
\qquad
B=\partial_H\mathcal M^{(-)}.
\label{eq:DB_from_Mminus_IV}
\end{equation}

To express this response in terms of the Lagrangian invariants, we expand
$\rho_-$ around the Maxwell limit. We define
\begin{equation}
s:=\sqrt{E^2+H^2},
\qquad
s>0,
\label{eq:s_def_IV}
\end{equation}
so that the expansion is regular in the nondegenerate electromagnetic
sector. From the definition of $\rho_-$,
\begin{equation}
\rho_-^2
=
(E+\beta q)^2+(H-\beta p)^2
=
s^2+2\beta(Eq-Hp)+\beta^2Q_c^2.
\label{eq:rho_minus_square_expand_Qc_IV}
\end{equation}
Equivalently,
\begin{equation}
\rho_-
=
s
\left[
1+
\frac{2\beta(Eq-Hp)+\beta^2Q_c^2}{s^2}
\right]^{1/2}.
\label{eq:rho_minus_before_taylor_IV}
\end{equation}
We now use the Taylor expansion
\begin{equation}
\begin{aligned}
(1+z)^{1/2}
&=
1+\frac{z}{2}+\mathcal O(z^2),
\\
z&:=\frac{2\beta(Eq-Hp)+\beta^2Q_c^2}{s^2} .
\end{aligned}
\label{eq:taylor_sqrt_IV}
\end{equation}
Since $z=\mathcal O(\beta)$, the term $z^2$ contributes only at
$\mathcal O(\beta^2)$. Hence
\begin{equation}
\rho_-
=
s+\frac{\beta(Eq-Hp)}{s}
+\mathcal O(\beta^2).
\label{eq:rho_minus_explicit_IV}
\end{equation}
Equivalently,
\begin{equation}
\frac{1}{\rho_-}
=
\frac{1}{s}
-
\beta\,\frac{Eq-Hp}{s^3}
+\mathcal O(\beta^2).
\label{eq:rho_minus_inverse_expansion_IV}
\end{equation}
In the inverse chart \eqref{eq:DB_inverse_local_exact_IV}, this quantity is
multiplied by $\beta$. Therefore, to first order it is enough to keep
\begin{equation}
\frac{\beta}{\rho_-}
=
\frac{\beta}{s}
+
\mathcal O(\beta^2).
\label{eq:beta_rhominus_inverse_IV}
\end{equation}
Thus
\begin{equation}
D
=
E-\beta Q_c\,\frac{E}{s}
+\mathcal O(\beta^2),
\qquad
B
=
H-\beta Q_c\,\frac{H}{s}
+\mathcal O(\beta^2).
\label{eq:DB_first_order_IV}
\end{equation}
From the second relation, $H=B+\mathcal O(\beta)$. Consequently, replacing
$H$ by $B$ inside $s$ changes the denominator only at order $\beta$;
since that denominator already multiplies $\beta$, the induced error is of
order $\beta^2$. We therefore introduce
\begin{equation}
s_0:=\sqrt{E^2+B^2},
\label{eq:s0_def_IV}
\end{equation}
and write
\begin{equation}
\begin{aligned}
D
&=
E\left(1-\frac{\beta Q_c}{s_0}\right)
+\mathcal O(\beta^2),
\\
H
&=
B\left(1+\frac{\beta Q_c}{s_0}\right)
+\mathcal O(\beta^2) .
\end{aligned}
\label{eq:DH_first_order_IV}
\end{equation}

With the conventions fixed in \eqref{eq:FGdef}--\eqref{eq:FGEB},
\begin{equation}
\mathcal F
=
\frac12(B^2-E^2),
\qquad
\mathcal G
=
-EB.
\label{eq:calFcalG_EB_IV}
\end{equation}
For a local Lagrangian NLED $L=L(\mathcal F,\mathcal G)$, the constitutive
relations in the principal frame are
\cite{Plebanski1970,Boillat1970}
\begin{equation}
D=L_{\mathcal F}E+L_{\mathcal G}B,
\qquad
H=L_{\mathcal F}B-L_{\mathcal G}E,
\label{eq:standard_constitutive_relations_IV}
\end{equation}
where
\begin{equation}
L_{\mathcal F}:=\frac{\partial L}{\partial\mathcal F},
\qquad
L_{\mathcal G}:=\frac{\partial L}{\partial\mathcal G}.
\label{eq:LF_LG_def_IV}
\end{equation}
With these conventions, the Maxwell limit corresponds to $L=\mathcal F$,
so that $L_{\mathcal F}=1$, $L_{\mathcal G}=0$, and one immediately
recovers $D=E$, $H=B$.

Solving \eqref{eq:standard_constitutive_relations_IV} for the first
derivatives gives
\begin{equation}
L_{\mathcal F}
=
\frac{ED+BH}{E^2+B^2},
\qquad
L_{\mathcal G}
=
\frac{BD-EH}{E^2+B^2}.
\label{eq:LF_LG_solution_IV}
\end{equation}
Substituting \eqref{eq:DH_first_order_IV}, one obtains
\begin{equation}
L_{\mathcal F}
=
1+\beta Q_c\,\frac{B^2-E^2}{(E^2+B^2)^{3/2}}
+\mathcal O(\beta^2),
\label{eq:LF_EB_IV}
\end{equation}
\begin{equation}
L_{\mathcal G}
=
-2\beta Q_c\,\frac{EB}{(E^2+B^2)^{3/2}}
+\mathcal O(\beta^2).
\label{eq:LG_EB_IV}
\end{equation}

We now define
\begin{equation}
\mathcal U:=\mathcal F^2+\mathcal G^2.
\label{eq:U_def_IV}
\end{equation}
Using \eqref{eq:calFcalG_EB_IV},
\begin{equation}
\mathcal U
=
\frac14(E^2+B^2)^2,
\qquad
E^2+B^2=2\sqrt{\mathcal U}.
\label{eq:E2B2_U_IV}
\end{equation}
Therefore, the derivatives above can be written as
\begin{equation}
\begin{aligned}
L_{\mathcal F}
&=
1+\frac{\beta Q_c}{\sqrt2}
\frac{\mathcal F}{\mathcal U^{3/4}}
+\mathcal O(\beta^2),
\\
L_{\mathcal G}
&=
\frac{\beta Q_c}{\sqrt2}
\frac{\mathcal G}{\mathcal U^{3/4}}
+\mathcal O(\beta^2) .
\end{aligned}
\label{eq:LF_LG_U_IV}
\end{equation}
These expressions are integrable at the order retained. Indeed,
\begin{equation}
\partial_{\mathcal F}\mathcal U^{1/4}
=
\frac12\,\mathcal F\,\mathcal U^{-3/4},
\qquad
\partial_{\mathcal G}\mathcal U^{1/4}
=
\frac12\,\mathcal G\,\mathcal U^{-3/4}.
\label{eq:U_quarter_derivatives_IV}
\end{equation}
Thus, a local Lagrangian representative compatible with
\eqref{eq:LF_LG_U_IV} is
\begin{equation}
L_{\rm loc}(\mathcal F,\mathcal G)
=
\mathcal F
+
\sqrt2\,\beta Q_c\,\mathcal U^{1/4}
+
\mathcal O(\beta^2),
\qquad
\mathcal U=\mathcal F^2+\mathcal G^2.
\label{eq:Lloc_final_IV}
\end{equation}
The additive constant has been omitted. The chart is defined for
\begin{equation}
\mathcal U>0,
\label{eq:U_positive_domain_IV}
\end{equation}
since at $\mathcal U=0$ the local representative is not differentiable.
This degenerate set lies outside the nontrivial electromagnetic sector
considered here.

\subsection{Fresnel coefficients}
\label{subsec:fresnel_coefficients}

The local representative \eqref{eq:Lloc_final_IV} contains the constitutive
information needed to construct the Fresnel quartic. In a Lagrangian NLED,
this information enters through the first and second derivatives of
$L(\mathcal F,\mathcal G)$ with respect to the electromagnetic invariants
\cite{Boillat1970,ObukhovRubilar2002}. At the order retained,
\begin{equation}
L_{\mathcal F\mathcal F}
=
\frac{\beta Q_c}{2\sqrt2}
\frac{2\mathcal G^2-\mathcal F^2}{\mathcal U^{7/4}}
+
\mathcal O(\beta^2),
\label{eq:LFF_IV}
\end{equation}
\begin{equation}
L_{\mathcal F\mathcal G}
=
-\frac{3\beta Q_c}{2\sqrt2}
\frac{\mathcal F\mathcal G}{\mathcal U^{7/4}}
+
\mathcal O(\beta^2),
\label{eq:LFG_IV}
\end{equation}
\begin{equation}
L_{\mathcal G\mathcal G}
=
\frac{\beta Q_c}{2\sqrt2}
\frac{2\mathcal F^2-\mathcal G^2}{\mathcal U^{7/4}}
+
\mathcal O(\beta^2).
\label{eq:LGG_IV}
\end{equation}
The sign of $L_{\mathcal F\mathcal G}$ is kept explicit because it
contributes linearly to the Fresnel coefficients.

We use the notation of Obukhov--Rubilar for the covariant characteristic
equation \cite{ObukhovRubilar2002},
\begin{equation}
\begin{aligned}
k_1&=4L_{\mathcal F},
&
k_2&=8L_{\mathcal F\mathcal F},
\\
k_3&=k_4=8L_{\mathcal F\mathcal G},
&
k_5&=8L_{\mathcal G\mathcal G} .
\end{aligned}
\label{eq:k12345_def_IV}
\end{equation}
Then
\begin{equation}
k_1
=
4
+
2\sqrt2\,\beta Q_c
\frac{\mathcal F}{\mathcal U^{3/4}}
+
\mathcal O(\beta^2),
\label{eq:k1_final_IV}
\end{equation}
\begin{equation}
k_2
=
2\sqrt2\,\beta Q_c
\frac{2\mathcal G^2-\mathcal F^2}{\mathcal U^{7/4}}
+
\mathcal O(\beta^2),
\label{eq:k2_final_IV}
\end{equation}
\begin{equation}
k_3=k_4
=
-6\sqrt2\,\beta Q_c
\frac{\mathcal F\mathcal G}{\mathcal U^{7/4}}
+
\mathcal O(\beta^2),
\label{eq:k3k4_final_IV}
\end{equation}
\begin{equation}
k_5
=
2\sqrt2\,\beta Q_c
\frac{2\mathcal F^2-\mathcal G^2}{\mathcal U^{7/4}}
+
\mathcal O(\beta^2).
\label{eq:k5_final_IV}
\end{equation}

With
\begin{equation}
I_1:=F_{\mu\nu}F^{\mu\nu}=4\mathcal F,
\qquad
I_2:=F_{\mu\nu}{}^\star F^{\mu\nu}=4\mathcal G,
\label{eq:I1I2_def_IV}
\end{equation}
the combinations that control the factorization of the quartic are
\cite{ObukhovRubilar2002}
\begin{equation}
X
=
k_1^2
+\frac{k_1}{2}(k_3+k_4)I_2
-k_1k_5I_1
+\frac14(k_3k_4-k_2k_5)I_2^2,
\label{eq:X_def_IV}
\end{equation}
\begin{equation}
Y
=
k_1(k_2+k_5)
+
(k_3k_4-k_2k_5)I_1,
\label{eq:Y_def_IV}
\end{equation}
\begin{equation}
Z
=
4(k_2k_5-k_3k_4).
\label{eq:Z_def_IV}
\end{equation}
For $X$ and $Y$ it is sufficient to retain terms through first order in
$\beta$. However, the discriminant involves the combination
$Y^2-XZ$. Since $Y=\mathcal O(\beta)$, the term $Y^2$ is of order
$\beta^2$; consequently, the leading $\mathcal O(\beta^2)$ contribution
to $Z$ must also be kept. Substituting
\eqref{eq:k1_final_IV}--\eqref{eq:k5_final_IV}, one obtains
\begin{equation}
X
=
16
-48\sqrt2\,\beta Q_c
\frac{\mathcal F}{\mathcal U^{3/4}}
+\mathcal O(\beta^2),
\label{eq:X_final_IV}
\end{equation}
\begin{equation}
Y
=
8\sqrt2\,\beta Q_c
\frac{1}{\mathcal U^{3/4}}
+\mathcal O(\beta^2).
\label{eq:Y_final_IV}
\end{equation}
Although $Z$ starts at order $\beta^2$, it must be retained in the
discriminant. Indeed,
\begin{equation}
Z
=
-64\,\beta^2Q_c^2\,\mathcal U^{-3/2}
+
\mathcal O(\beta^3).
\label{eq:Z_quadratic_IV}
\end{equation}
The reason for keeping this term, despite its starting at order $\beta^2$,
is that the separation of the two optical roots depends on the radical
\begin{equation}
\sqrt{Y^2-XZ}.
\label{eq:fresnel_radical_IV}
\end{equation}
Since $Y=\mathcal O(\beta)$ and $X=16+\mathcal O(\beta)$, both $Y^2$
and $XZ$ contribute to the radical at order $\beta^2$. Indeed,
\begin{equation}
Y^2
=
128\,\beta^2Q_c^2\,\mathcal U^{-3/2}
+
\mathcal O(\beta^3),
\label{eq:Y_squared_IV}
\end{equation}
whereas
\begin{equation}
-XZ
=
1024\,\beta^2Q_c^2\,\mathcal U^{-3/2}
+
\mathcal O(\beta^3).
\label{eq:minus_XZ_IV}
\end{equation}
Therefore,
\begin{equation}
Y^2-XZ
=
1152\,\beta^2Q_c^2\,\mathcal U^{-3/2}
+
\mathcal O(\beta^3),
\label{eq:discriminant_expansion_IV}
\end{equation}
and consequently
\begin{equation}
\sqrt{Y^2-XZ}
=
24\sqrt2\,|\beta|\,Q_c\,\mathcal U^{-3/4}
+
\mathcal O(\beta^2).
\label{eq:fresnel_radical_expanded_IV}
\end{equation}
We define
\begin{equation}
\Delta_{\!F}:=\sqrt{Y^2-XZ},
\qquad
\Delta_{\!F}\ge0.
\label{eq:DeltaF_def_IV}
\end{equation}
At leading order,
\begin{equation}
\Delta_{\!F}
=
24\sqrt2\,|\beta|\,Q_c\,\mathcal U^{-3/4}
+
\mathcal O(\beta^2).
\label{eq:DeltaF_leading_IV}
\end{equation}
The square root is defined to be non-negative. Therefore, changing the sign
of $\beta$ does not change the two optical cones themselves, but it can
interchange the labels attached to the two roots. In what follows, the
labels $+$ and $-$ are understood within a fixed perturbative choice of
the sign of $\beta$.

\subsection{Quartic factorization and optical metrics}
\label{subsec:fresnel_factorization}

With the coefficients above, the Fresnel quartic factorizes into two
quadratic forms for the wave covector $\ell_\mu$
\cite{ObukhovRubilar2002}. Denoting by
$\mathscr G_{\rm TR}^{\mu\nu\rho\sigma}$ the Tamm--Rubilar tensor
associated with the characteristic equation, one has
\begin{equation}
\begin{aligned}
&\mathscr G_{\rm TR}^{\mu\nu\rho\sigma}
\ell_\mu \ell_\nu \ell_\rho \ell_\sigma
\\
&\qquad =
-\frac{k_1}{8X}
\left(g_+^{\alpha\beta}\ell_\alpha\ell_\beta\right)
\left(g_-^{\gamma\delta}\ell_\gamma\ell_\delta\right)
=0 .
\end{aligned}
\label{eq:fresnel_factorized_IV}
\end{equation}
Thus, propagation is governed by one of the two quadratic conditions
\begin{equation}
g_\pm^{\mu\nu}\ell_\mu\ell_\nu=0 .
\label{eq:two_optical_cones_IV}
\end{equation}
Each condition defines a conformal class of contravariant optical metrics,
\begin{equation}
g_\pm^{\mu\nu}
=
Xg^{\mu\nu}
+
\left(Y\pm\Delta_{\!F}\right)t^{\mu\nu},
\label{eq:gpm_general_IV}
\end{equation}
where
\begin{equation}
t^{\mu\nu}=F^\mu{}_{\alpha}F^{\alpha\nu}.
\label{eq:tmunu_optical_IV}
\end{equation}
This is the quadratic tensor introduced in
\eqref{eq:tmunu_background_def}. Its appearance shows that the optical
correction follows the directions algebraically selected by the background
electromagnetic field.

Since unparametrized null trajectories are invariant under regular conformal
rescalings, we can extract the global factor $X$. Indeed,
\begin{equation}
g_\pm^{\mu\nu}
=
X
\left[
g^{\mu\nu}
+
\frac{Y\pm\Delta_{\!F}}{X}\,
t^{\mu\nu}
\right].
\label{eq:gpm_factor_X_IV}
\end{equation}
In the perturbative regime considered here,
\begin{equation}
X=16+\mathcal O(\beta),
\label{eq:X_positive_perturbative_IV}
\end{equation}
so that $X$ remains nonzero for sufficiently small $|\beta|$ within the
domain of validity. The conformal factor can therefore be removed without
changing the null directions, and we work with the regular representative
\begin{equation}
\tilde g_\pm^{\mu\nu}
=
g^{\mu\nu}
+
\nu_\pm\,t^{\mu\nu}
+
\mathcal O(\beta^2),
\label{eq:gpm_conformal_IV}
\end{equation}
with
\begin{equation}
\nu_\pm
:=
\frac{Y\pm\Delta_{\!F}}{X}.
\label{eq:nupm_def_IV}
\end{equation}

Since $Y=\mathcal O(\beta)$, $\Delta_{\!F}=\mathcal O(\beta)$, and
$X=16+\mathcal O(\beta)$, one has
\begin{equation}
\nu_\pm=\mathcal O(\beta).
\label{eq:nupm_order_IV}
\end{equation}
On the other hand, the exact fields admit the expansion
\begin{equation}
E=E_0+\mathcal O(\beta),
\qquad
B=B_0+\mathcal O(\beta).
\label{eq:fields_linear_limit_IV}
\end{equation}
Therefore,
\begin{equation}
\nu_\pm E^2
=
\nu_\pm E_0^2
+
\mathcal O(\beta^2),
\qquad
\nu_\pm B^2
=
\nu_\pm B_0^2
+
\mathcal O(\beta^2).
\label{eq:nupm_fields_order_IV}
\end{equation}
Thus, at the order retained, it is enough to evaluate the fields in the
linear limit. From \eqref{eq:Exy} and \eqref{eq:Bxy}, one obtains
\begin{equation}
E_0(r,\theta)
=
\frac{
q(a^2\cos^2\theta-r^2)+2par\cos\theta
}{\Sigma^2},
\label{eq:E0_def_IV}
\end{equation}
\begin{equation}
B_0(r,\theta)
=
\frac{
p(r^2-a^2\cos^2\theta)+2qar\cos\theta
}{\Sigma^2}.
\label{eq:B0_def_IV}
\end{equation}
Then
\begin{equation}
\mathcal F_0=\frac12(B_0^2-E_0^2),
\qquad
\mathcal G_0=-E_0B_0,
\qquad
\mathcal U_0=\mathcal F_0^2+\mathcal G_0^2.
\label{eq:F0G0U0_def_IV}
\end{equation}
In the linear limit one has $E_0=\mathcal A/\Sigma^2$ and
$B_0=\mathcal C/\Sigma^2$. Therefore,
\begin{equation}
E_0^2+B_0^2
=
\frac{\mathcal A^2+\mathcal C^2}{\Sigma^4}
=
\frac{Q_c^2}{\Sigma^2},
\label{eq:E0B0_closed_derivation_IV}
\end{equation}
where we have used \eqref{eq:AC_norm_identity}. Hence
\begin{equation}
\mathcal U_0
=
\frac14(E_0^2+B_0^2)^2
=
\frac{Q_c^4}{4\Sigma^4}.
\label{eq:U0_closed_IV}
\end{equation}

For a perturbative choice with $\beta>0$, one has
\begin{equation}
\begin{aligned}
Y
&=
8\sqrt2\beta Q_c\mathcal U_0^{-3/4}
+\mathcal O(\beta^2),
\\
\Delta_{\!F}
&=
24\sqrt2\beta Q_c\mathcal U_0^{-3/4}
+\mathcal O(\beta^2) .
\end{aligned}
\label{eq:Y_Delta_beta_positive_IV}
\end{equation}
and $X=16+\mathcal O(\beta)$. Therefore,
\begin{equation}
\begin{aligned}
\nu_+
&=
\frac{8+24}{16}
\sqrt2\beta Q_c\mathcal U_0^{-3/4}
+\mathcal O(\beta^2),
\\
\nu_-
&=
\frac{8-24}{16}
\sqrt2\beta Q_c\mathcal U_0^{-3/4}
+\mathcal O(\beta^2) .
\end{aligned}
\label{eq:nupm_intermediate_IV}
\end{equation}
That is,
\begin{equation}
\begin{aligned}
\nu_+
&=
2\sqrt2\beta Q_c\mathcal U_0^{-3/4}
+\mathcal O(\beta^2),
\\
\nu_-
&=
-\sqrt2\beta Q_c\mathcal U_0^{-3/4}
+\mathcal O(\beta^2) .
\end{aligned}
\label{eq:nupm_explicit_U0_IV}
\end{equation}
Using \eqref{eq:U0_closed_IV},
\begin{equation}
\mathcal U_0^{-3/4}
=
\left(\frac{Q_c^4}{4\Sigma^4}\right)^{-3/4}
=
2^{3/2}\frac{\Sigma^3}{Q_c^3},
\label{eq:U0_minus_three_quarters_IV}
\end{equation}
and therefore
\begin{equation}
\nu_+
=
8\beta\,\frac{\Sigma^3}{Q_c^2}
+
\mathcal O(\beta^2),
\qquad
\nu_-
=
-4\beta\,\frac{\Sigma^3}{Q_c^2}
+
\mathcal O(\beta^2).
\label{eq:nupm_explicit_background_IV}
\end{equation}
If the sign of $\beta$ is changed, the convention
$\Delta_{\!F}\ge0$ interchanges the labels of the two roots. Therefore, the
general expressions remain those given in \eqref{eq:nupm_def_IV}, and the
labels $+$ and $-$ are always understood within a fixed-sign
perturbative choice. We do not identify any branch as ordinary or
extraordinary; we distinguish only the two optical roots of the Fresnel
quartic.

\subsection{Tetrad form and optical domain}
\label{subsec:tetrad_optics}

The geometrical interpretation of the two Fresnel roots becomes transparent
in the principal tetrad. In that frame, the background metric satisfies
\begin{equation}
g^{\hat a\hat b}
=
\mathrm{diag}(-1,1,1,1),
\label{eq:g_tetrad_background_IV}
\end{equation}
whereas the quadratic tensor of the aligned field, introduced in
\eqref{eq:tmunu_background_def}, takes the form
\begin{equation}
t^{\hat a\hat b}
=
\mathrm{diag}
\bigl(
-E^2,\,
+E^2,\,
-B^2,\,
-B^2
\bigr).
\label{eq:tetrad_t_corrected_IV}
\end{equation}
Substituting \eqref{eq:tetrad_t_corrected_IV} into
\eqref{eq:gpm_conformal_IV}, one obtains
\begin{equation}
\tilde g_\pm^{\hat0\hat0}
=
-\bigl(1+\nu_\pm E^2\bigr)
+
\mathcal O(\beta^2),
\qquad
\tilde g_\pm^{\hat1\hat1}
=
1+\nu_\pm E^2
+
\mathcal O(\beta^2),
\label{eq:g0011_tetrad_derivation_IV}
\end{equation}
and
\begin{equation}
\tilde g_\pm^{\hat2\hat2}
=
1-\nu_\pm B^2
+
\mathcal O(\beta^2),
\qquad
\tilde g_\pm^{\hat3\hat3}
=
1-\nu_\pm B^2
+
\mathcal O(\beta^2).
\label{eq:g2233_tetrad_derivation_IV}
\end{equation}
The background electric field therefore deforms the temporal--radial block,
whereas the magnetic field deforms the angular block.

Using the perturbative counting established in
\eqref{eq:nupm_fields_order_IV}, the optical corrections can be evaluated
with the fields in the linear limit. We define
\begin{equation}
\begin{aligned}
\Upsilon_\pm(r,\theta)
&:=
1+\nu_\pm(r,\theta)E_0(r,\theta)^2,
\\
\Phi_\pm(r,\theta)
&:=
1-\nu_\pm(r,\theta)B_0(r,\theta)^2 .
\end{aligned}
\label{eq:upsilon_phi_def_IV}
\end{equation}
With this notation, the conformal representative of the optical metrics
takes the diagonal form
\begin{equation}
\tilde g_\pm^{\hat a\hat b}
=
\mathrm{diag}
\bigl(
-\Upsilon_\pm,\,
+\Upsilon_\pm,\,
+\Phi_\pm,\,
+\Phi_\pm
\bigr)
+
\mathcal O(\beta^2).
\label{eq:gpm_tetrad_observables_IV}
\end{equation}
The corresponding covariant metric is obtained by inverting this diagonal
matrix:
\begin{equation}
\tilde g_{\pm\,\hat a\hat b}
=
\mathrm{diag}
\left(
-\frac{1}{\Upsilon_\pm},\,
\frac{1}{\Upsilon_\pm},\,
\frac{1}{\Phi_\pm},\,
\frac{1}{\Phi_\pm}
\right)
+
\mathcal O(\beta^2).
\label{eq:gpm_cov_tetrad_IV}
\end{equation}
Since $\Upsilon_\pm=1+\mathcal O(\beta)$ and
$\Phi_\pm=1+\mathcal O(\beta)$, this expression is equivalent, at the
order retained, to expanding the inverses in series.

In coordinates, the contravariant form is reconstructed with the dual tetrad
basis:
\begin{equation}
\tilde g_\pm^{\mu\nu}
=
-\Upsilon_\pm\,e_{\hat0}^{\mu}e_{\hat0}^{\nu}
+
\Upsilon_\pm\,e_{\hat1}^{\mu}e_{\hat1}^{\nu}
+
\Phi_\pm\,e_{\hat2}^{\mu}e_{\hat2}^{\nu}
+
\Phi_\pm\,e_{\hat3}^{\mu}e_{\hat3}^{\nu}
+
\mathcal O(\beta^2).
\label{eq:gpm_coord_from_tetrad_IV}
\end{equation}
If $\ell_{\hat a}=(-\omega,K_1,K_2,K_3)$ is the wave covector measured in
the principal tetrad, the null condition for each branch is
\begin{equation}
-\Upsilon_\pm\,\omega^2
+
\Upsilon_\pm\,K_1^2
+
\Phi_\pm(K_2^2+K_3^2)
=
0
+
\mathcal O(\beta^2).
\label{eq:pm_cone_IV}
\end{equation}
The difference between $\nu_+$ and $\nu_-$ is therefore translated into
two distinct local dispersion relations. This is the tetrad manifestation of
the birefringent splitting of the propagation cones.

For each optical representative to preserve Lorentzian signature in the
domain of interest, we require
\begin{equation}
\Upsilon_\pm(r,\theta)>0,
\qquad
\Phi_\pm(r,\theta)>0.
\label{eq:optical_signature_conditions_IV}
\end{equation}
These inequalities are domain conditions, not dynamical equations: they
ensure that the temporal--radial and angular blocks do not change sign in the
region where the optical representative is used. The perturbative
approximation also requires the corrections to unity to remain small.
Therefore, we introduce the control parameter
\begin{equation}
\epsilon_{\rm pert}
:=
\max_\pm
\sup_{(r,\theta)\in\mathcal D}
\max
\left\{
|\nu_\pm E_0^2|,
|\nu_\pm B_0^2|
\right\}
\ll1.
\label{eq:epsilon_pert_IV}
\end{equation}
For the critical curves we will consider a domain
\begin{equation}
\mathcal D
=
\{(r,\theta):r_+<r<r_L,\ 0<\theta<\pi\},
\label{eq:optical_domain_D_IV}
\end{equation}
where $r_+$ is the outer root of $\Delta(r)=0$ and $r_L$ fixes the
radial scale up to which the optical description is controlled. No uniform
validity up to infinity is assumed; at large radii, perturbative control is
determined by \eqref{eq:epsilon_pert_IV}.

The location of the radial null surface is read directly from
\eqref{eq:gpm_coord_from_tetrad_IV}. Since
\begin{equation}
e_{\hat1}^{r}
=
\sqrt{\frac{\Delta}{\Sigma}},
\label{eq:e1r_component_IV}
\end{equation}
one obtains
\begin{equation}
\tilde g_\pm^{rr}
=
\Upsilon_\pm\,\frac{\Delta}{\Sigma}
+
\mathcal O(\beta^2).
\label{eq:grr_optical_IV}
\end{equation}
The optical correction multiplies $g^{rr}$ by a regular factor, but it
does not shift its zeros at the order considered. As long as
$\Upsilon_\pm$ is finite and positive, the surface
\begin{equation}
\Delta(r)=0
\label{eq:Delta_horizon_recalled_IV}
\end{equation}
remains null for the optical representatives. Therefore, the outer root
$r_+$ remains the natural optical capture surface within the perturbative
regime considered.

In summary, the Fresnel construction produces two conformal classes of
optical metrics,
\begin{equation}
\tilde g_\pm^{\hat a\hat b}
=
\mathrm{diag}
\bigl(
-\Upsilon_\pm,\,
+\Upsilon_\pm,\,
+\Phi_\pm,\,
+\Phi_\pm
\bigr)
+
\mathcal O(\beta^2),
\label{eq:optical_metrics_summary_IV}
\end{equation}
with
\begin{equation}
\begin{aligned}
\Upsilon_\pm&=1+\nu_\pm E_0^2,
&
\Phi_\pm&=1-\nu_\pm B_0^2,
\\
\nu_\pm&=\frac{Y\pm\sqrt{Y^2-XZ}}{X} .
\end{aligned}
\label{eq:optical_coefficients_summary_IV}
\end{equation}
The result preserves the block structure of the principal tetrad: the
temporal--radial sector and the angular sector are deformed by different
factors, and the two Fresnel roots encode the local birefringence of the
branch.


\section{Separability of the optical branches}
\label{sec:HJ}

The two optical metrics obtained above give the characteristic cones for the
two Fresnel branches.  The next question is dynamical.  In a rotating
spacetime the critical contour is not fixed by one radius only; it depends on
the possibility of separating the Hamilton--Jacobi equation and constructing
radial and angular potentials.  For the usual Kerr family this is the role of
Carter separability \cite{Carter1968HJ,Carter1968Global,Chandrasekhar1983}.
Here we ask whether the same type of structure survives for the optical
metrics generated by the NLED response.

The answer, within the perturbative order used in this paper, is yes.  The
conformal factors that appear in the two optical metrics keep a radial--angular
structure that permits separation.  This gives one separation constant for
each optical branch and allows us to define branch-dependent critical
families.

\subsection{Hamilton--Jacobi equation for the optical metrics}
\label{subsec:HJ_pm}

For each optical branch we use the null Hamiltonian
\begin{equation}
\mathcal H_\pm(x,p)
=
\frac12\,\tilde g_\pm^{\mu\nu}(x)p_\mu p_\nu
=
0,
\qquad
p_\mu=\partial_\mu S_\pm .
\label{eq:Hamiltonian_pm_HJ}
\end{equation}
The difference with the usual Kerr-Newman case lies not in the Hamiltonian form, but in the cone that defines the null condition: each Fresnel root has its own optical representative $\tilde g_\pm^{\mu\nu}$.

We work with the dual frame of the principal coframe introduced in
Sec.~\ref{sec:background},
\begin{eqnarray}
    \label{eq:dual_frame01_HJ}
    e_{\hat 0}
&=&
\frac{r^2+a^2}{\sqrt{\Delta\Sigma}}\,\partial_t
+
\frac{a}{\sqrt{\Delta\Sigma}}\,\partial_\phi,\\
\label{eq:dual_frame01_HJ2}
e_{\hat 1}
&=&
\sqrt{\frac{\Delta}{\Sigma}}\,\partial_r,\\
\label{eq:dual_frame23_HJ}
e_{\hat 2}
&=&
\frac{1}{\sqrt{\Sigma}}\,\partial_\theta,\\
\label{eq:dual_frame23_HJ2}
e_{\hat 3}
&=&
\frac{a\sin\theta}{\sqrt{\Sigma}}\,\partial_t
+
\frac{1}{\sqrt{\Sigma}\sin\theta}\,\partial_\phi .
\end{eqnarray}
In this frame, the optical metrics preserve the block structure obtained in
\eqref{eq:gpm_tetrad_observables_IV},
\begin{equation}
\tilde g_\pm^{\hat a\hat b}
=
\mathrm{diag}
\bigl(
-\Upsilon_\pm,\,
+\Upsilon_\pm,\,
+\Phi_\pm,\,
+\Phi_\pm
\bigr)
+
\mathcal O(\beta^2).
\label{eq:gpm_tetrad_recalled_HJ}
\end{equation}
The factor $\Upsilon_\pm$ deforms the temporal--radial plane, whereas
$\Phi_\pm$ deforms the angular block.

Stationarity and axisymmetry give the conserved quantities associated with
$\partial_t$ and $\partial_\phi$,
\begin{equation}
\mathcal E:=-p_t,
\qquad
L:=p_\phi .
\label{eq:EL_constants_HJ}
\end{equation}
We therefore write
\begin{equation}
S_\pm
=
-\mathcal E\,t
+
L\,\phi
+
W_\pm(r,\theta),
\label{eq:S_pm_ansatz_HJ}
\end{equation}
and introduce the combinations
\begin{equation}
T(r):=(r^2+a^2)\mathcal E-aL,
\qquad
J(\theta):=L-a\mathcal E\sin^2\theta .
\label{eq:TJ_def_HJ}
\end{equation}
By definition, the tetrad components of the covector are
$p_{\hat a}=e_{\hat a}{}^\mu p_\mu$. Using $p_t=-\mathcal E$ and
$p_\phi=L$, the combinations above give
\begin{equation}
p_{\hat0}
=
-\frac{T(r)}{\sqrt{\Delta\Sigma}},
\qquad
p_{\hat1}
=
\sqrt{\frac{\Delta}{\Sigma}}\,
\partial_r W_\pm,
\label{eq:p0p1_pm_HJ}
\end{equation}
\begin{equation}
p_{\hat2}
=
\frac{\partial_\theta W_\pm}{\sqrt{\Sigma}},
\qquad
p_{\hat3}
=
\frac{J(\theta)}{\sqrt{\Sigma}\sin\theta}.
\label{eq:p2p3_pm_HJ}
\end{equation}
Substituting these components into
$\tilde g_\pm^{\hat a\hat b}p_{\hat a}p_{\hat b}=0$ and multiplying by
$\Sigma$, one obtains
\begin{equation}
\begin{aligned}
&\Upsilon_\pm
\left[
\Delta(r)\bigl(\partial_r W_\pm\bigr)^2
-
\frac{T(r)^2}{\Delta(r)}
\right]
\\
&\quad+
\Phi_\pm
\left[
\bigl(\partial_\theta W_\pm\bigr)^2
+
\frac{J(\theta)^2}{\sin^2\theta}
\right]
=\mathcal O(\beta^2) .
\end{aligned}
\label{eq:HJ_pm_block_HJ}
\end{equation}
The equation retains the radial--angular organization characteristic of
Carter's problem, although effective separability now depends on the
structure of the optical factors that weight the two blocks.

\subsection{Conformal ratio and separability}
\label{subsec:HJ_coefficients}

Multiplying the null Hamilton--Jacobi equation by a regular and nonzero
conformal factor does not change the set of null directions. Therefore, after
\eqref{eq:HJ_pm_block_HJ}, and in the perturbative domain where
$\Phi_\pm=1+\mathcal O(\beta)$ remains nonzero, only the ratio between the
two optical factors can affect radial--angular separation:
\begin{equation}
\frac{\Upsilon_\pm}{\Phi_\pm}.
\label{eq:conformal_ratio_relevant_HJ}
\end{equation}
This observation prevents us from assigning dynamical meaning to a
particular conformal representative: what matters is the relation between
the deformation of the temporal--radial block and that of the angular block.

To evaluate this ratio, it is enough to use the fields in the linear limit,
because $\nu_\pm=\mathcal O(\beta)$. We write
\begin{eqnarray}
  \label{eq:E0B0_def_HJ}
  E_0(r,\theta)&=\frac{N_E(r,\theta)}{\Sigma^2},\\
  \label{eq:E0B0_def_HJ2}
B_0(r,\theta)&=\frac{N_B(r,\theta)}{\Sigma^2}, 
\end{eqnarray}where
\begin{eqnarray}\label{eq:NE_NB_def_HJ} 
   N_E={}&q(a^2\cos^2\theta-r^2)+2par\cos\theta,\\
N_B={}&p(r^2-a^2\cos^2\theta)+2qar\cos\theta .
\label{eq:NE_NB_def_HJ2} 
\end{eqnarray}

In the linear limit, these quantities coincide with the combinations
$\mathcal A$ and $\mathcal C$ introduced in \eqref{eq:AC_def}. Indeed,
$N_E=\mathcal A$ and $N_B=\mathcal C$, with $x=a\cos\theta$ and
$y=r$. Therefore, using \eqref{eq:AC_norm_identity},
\begin{equation}
N_E^2+N_B^2
=
Q_c^2\Sigma^2,
\qquad
Q_c^2=p^2+q^2,
\label{eq:NE_NB_identity_HJ}
\end{equation}
and
\begin{equation}
E_0^2+B_0^2
=
\frac{Q_c^2}{\Sigma^2}.
\label{eq:E0B0_sum_HJ}
\end{equation}

From the factorization of the quartic in Sec.~\ref{sec:opticalmetrics}, each
optical branch is controlled by
\begin{equation}
\nu_\pm
=
\frac{Y\pm\Delta_{\!F}}{X}.
\label{eq:nupm_from_fresnel_HJ}
\end{equation}
Substituting \eqref{eq:nupm_explicit_background_IV}, both cases can be
written, on the charged branch $Q_c>0$, as
\begin{equation}
\nu_\pm
=
\beta\,\alpha_\pm\,\frac{\Sigma^3}{Q_c^2}
+
\mathcal O(\beta^2).
\label{eq:nupm_alpha_HJ}
\end{equation}
For the convention $\Delta_{\!F}\ge0$ and a perturbative choice with
$\beta>0$,
\begin{equation}
\alpha_+=8,
\qquad
\alpha_-=-4.
\label{eq:alpha_values_HJ}
\end{equation}
If the sign of $\beta$ is reversed, the non-negative-root convention may
interchange the labels of the two solutions; hence the labels $+$ and
$-$ are always understood within a fixed-sign perturbative choice.

Using
\begin{equation}
\Upsilon_\pm=1+\nu_\pm E_0^2,
\qquad
\Phi_\pm=1-\nu_\pm B_0^2,
\label{eq:UpsilonPhi_recalled_HJ}
\end{equation}
one obtains, to first order,
\begin{equation}
\frac{\Upsilon_\pm}{\Phi_\pm}
=
\bigl(1+\nu_\pm E_0^2\bigr)
\bigl(1+\nu_\pm B_0^2\bigr)
+
\mathcal O(\beta^2).
\label{eq:ratio_first_step_HJ}
\end{equation}
Since $\nu_\pm=\mathcal O(\beta)$, this gives
\begin{equation}
\frac{\Upsilon_\pm}{\Phi_\pm}
=
1+\nu_\pm(E_0^2+B_0^2)
+
\mathcal O(\beta^2).
\label{eq:ratio_second_step_HJ}
\end{equation}
With \eqref{eq:E0B0_sum_HJ} and \eqref{eq:nupm_alpha_HJ},
\begin{equation}
\frac{\Upsilon_\pm}{\Phi_\pm}
=
1+\beta\alpha_\pm\Sigma
+
\mathcal O(\beta^2).
\label{eq:ratio_UpsilonPhi_sigma_HJ}
\end{equation}
Finally, since
\begin{equation}
\Sigma=r^2+a^2\cos^2\theta,
\label{eq:Sigma_ru_HJ}
\end{equation}
the ratio admits the perturbative factorization
\begin{equation}
1+\beta\alpha_\pm\Sigma
=
\bigl(1+\beta\alpha_\pm r^2\bigr)
\bigl(1+\beta\alpha_\pm a^2\cos^2\theta\bigr)
+
\mathcal O(\beta^2).
\label{eq:ratio_factorization_HJ}
\end{equation}
Although $\Upsilon_\pm$ and $\Phi_\pm$ separately contain mixed
dependences, their conformal ratio preserves a radial--angular separable
structure at first order.

\subsection{Carter-type separation}
\label{subsec:separability_HJ}

We divide \eqref{eq:HJ_pm_block_HJ} by $\Phi_\pm$, which is legitimate
inside the optical domain where $\Phi_\pm>0$, and take
\begin{equation}
W_\pm(r,\theta)=S_r(r)+S_\theta(\theta).
\label{eq:additive_W_HJ}
\end{equation}
We define the blocks
\begin{equation}
\mathscr R(r)
:=
\Delta(r)\bigl(S_r'(r)\bigr)^2
-
\frac{T(r)^2}{\Delta(r)},
\label{eq:Rblock_HJ}
\end{equation}
\begin{equation}
\mathscr T(\theta)
:=
\bigl(S_\theta'(\theta)\bigr)^2
+
\frac{J(\theta)^2}{\sin^2\theta}.
\label{eq:Tblock_HJ}
\end{equation}
Then
\begin{equation}
\frac{\Upsilon_\pm}{\Phi_\pm}\,
\mathscr R(r)
+
\mathscr T(\theta)
=
0
+
\mathcal O(\beta^2).
\label{eq:HJ_divided_by_Phi}
\end{equation}
Substituting \eqref{eq:ratio_factorization_HJ},
\begin{equation}
\bigl(1+\beta\alpha_\pm r^2\bigr)
\bigl(1+\beta\alpha_\pm a^2\cos^2\theta\bigr)
\mathscr R(r)
+
\mathscr T(\theta)
=
0
+
\mathcal O(\beta^2).
\label{eq:HJ_before_angular_division_HJ}
\end{equation}
We now multiply by
$(1+\beta\alpha_\pm a^2\cos^2\theta)^{-1}$. To first order,
\begin{equation}
\frac{1}{1+\beta\alpha_\pm a^2\cos^2\theta}
=
1-\beta\alpha_\pm a^2\cos^2\theta
+
\mathcal O(\beta^2).
\label{eq:angular_inverse_factor_HJ}
\end{equation}
Therefore,
\begin{equation}
\bigl(1+\beta\alpha_\pm r^2\bigr)\,
\mathscr R(r)
+
\bigl(1-\beta\alpha_\pm a^2\cos^2\theta\bigr)\,
\mathscr T(\theta)
=
0
+
\mathcal O(\beta^2).
\label{eq:HJ_separable_form}
\end{equation}
The first contribution depends only on $r$, and the second only on
$\theta$. The two optical branches therefore admit Hamilton--Jacobi
separation at first order.

We introduce the separation constant $\mathcal K_\pm$ through
\begin{equation}
\bigl(1+\beta\alpha_\pm r^2\bigr)
\left[
\Delta(r)\bigl(S_r'(r)\bigr)^2
-
\frac{T(r)^2}{\Delta(r)}
\right]
=
-\mathcal K_\pm,
\label{eq:radial_separation_HJ}
\end{equation}
\begin{equation}
\bigl(1-\beta\alpha_\pm a^2\cos^2\theta\bigr)
\left[
\bigl(S_\theta'(\theta)\bigr)^2
+
\frac{J(\theta)^2}{\sin^2\theta}
\right]
=
\mathcal K_\pm.
\label{eq:angular_separation_HJ}
\end{equation}
Equivalently,
\begin{equation}
\Delta(r)\bigl(S_r'(r)\bigr)^2
=
\frac{T(r)^2}{\Delta(r)}
-
\frac{\mathcal K_\pm}{1+\beta\alpha_\pm r^2}
+
\mathcal O(\beta^2),
\label{eq:Sr_equation_HJ}
\end{equation}
\begin{equation}
\bigl(S_\theta'(\theta)\bigr)^2
=
\frac{\mathcal K_\pm}
{1-\beta\alpha_\pm a^2\cos^2\theta}
-
\frac{J(\theta)^2}{\sin^2\theta}
+
\mathcal O(\beta^2).
\label{eq:Stheta_equation_HJ}
\end{equation}
The constant $\mathcal K_\pm$ is the branch-dependent optical counterpart
of Carter's constant. In the limit $\beta\to0$, the two Fresnel roots
collapse to the same cone, and the equations above recover the usual null
separation of the Kerr--Newman sector \cite{Carter1968HJ,Chandrasekhar1983}.

The statement is perturbative: the optical representatives constructed at
first order are separable at that same order. Whether this structure
persists beyond the order considered depends on the $\mathcal O(\beta^2)$
corrections of the full optical reconstruction.

\subsection{Limits and physical interpretation}
\label{subsec:special_regimes_HJ}

The separated structure above has two control limits. In the Maxwell limit,
$\beta\to0$, the terms proportional to $\beta\alpha_\pm$ disappear. The
two Fresnel roots then collapse into the same null cone, and the separated
equations recover the usual null structure of the Kerr--Newman sector.

The second control is the nonrotating limit. If
\begin{equation}
a=0,
\qquad
\Sigma=r^2,
\qquad
J(\theta)=L,
\label{eq:nonrotating_limit_HJ}
\end{equation}
the angular equation loses all dependence on the optical branch and takes
the spherically symmetric form
\begin{equation}
\bigl(S_\theta'(\theta)\bigr)^2
+
\frac{L^2}{\sin^2\theta}
=
\mathcal K_\pm
+
\mathcal O(\beta^2).
\label{eq:nonrotating_angular_HJ}
\end{equation}
The first-order optical correction is then concentrated in the radial
sector, as expected when rotation no longer distinguishes preferred angular
directions.

Altogether, each optical Fresnel root admits a Carter-type separation at
first order and carries its own constant $\mathcal K_\pm$. The difference
between branches is encoded in the coefficients $\alpha_\pm$, which deform
the radial and angular factors differently. This separability is not imposed
as an additional hypothesis: at the perturbative order considered, it arises
from the combination of the Carter-type structure of the rotating geometry
and the two Fresnel roots of the NLED.


\section{Birefringent critical curves and local projection}
\label{sec:birefringent_critical_curves}

We now connect the separated optical dynamics with the image seen by a local
observer.  For each optical branch we impose the critical conditions on the
radial potential and obtain the corresponding constants of motion.  These
constants are then projected on the celestial sphere of an observer at finite
distance.  This finite-distance construction is important because the
Garc\'ia--D\'iaz branch is not asymptotically flat when $\beta\neq0$.

The output of the construction is a pair of critical contours.  We call them
$\Gamma_+$ and $\Gamma_-$.  They coincide in the Maxwell limit, and their
separation gives the geometrical birefringent signal.  The aim of this section
is to define this signal in a way that is independent of any emission model.

\subsection{Critical families of the optical branches}
\label{subsec:critical_photon_shells}

For each optical branch $s=\pm$, the separation obtained in
Sec.~\ref{sec:HJ} introduces the factors
\begin{equation}
A_s(r):=1+\beta\alpha_s r^2,
\qquad
B_s(\theta):=1-\beta\alpha_s a^2\cos^2\theta .
\label{eq:As_Bs_def}
\end{equation}
The coefficients $\alpha_s$ label the two Fresnel roots. With the
convention $\Delta_{\!F}\ge0$ and a perturbative choice with
$\beta>0$,
\begin{equation}
\alpha_+=8,
\qquad
\alpha_-=-4.
\label{eq:alpha_pm_VI}
\end{equation}
If the sign of $\beta$ is changed, the non-negative-root convention may
interchange the labels $+$ and $-$. Therefore, the expressions will be
kept in terms of $\alpha_s$.

Stationarity and axisymmetry conserve
\begin{equation}
\mathcal E:=-p_t,
\qquad
L:=p_\phi,
\label{eq:VI_EL_def}
\end{equation}
and the Carter-type separation introduces a constant $\mathcal K_s$ for
each branch. We use the dimensionless quantities
\begin{equation}
\xi_s:=\frac{L}{\mathcal E},
\qquad
k_s:=\frac{\mathcal K_s}{\mathcal E^2}.
\label{eq:xi_k_def_VI}
\end{equation}
The subscript $s$ does not indicate an energy or an angular momentum of a
different nature; it indicates that the critical values selected by the
optical dynamics depend on the branch. This is the birefringent version of
the critical parametrization associated with Carter separation in rotating
geometries \cite{Carter1968HJ,Chandrasekhar1983}.

From \eqref{eq:Sr_equation_HJ}--\eqref{eq:Stheta_equation_HJ}, the separated
potentials can be written as
\begin{equation}
\begin{aligned}
\mathcal R_s(r;\xi_s,k_s)
&=
P_s(r)^2-\frac{\Delta(r)}{A_s(r)}k_s,
\\
P_s(r)&:=r^2+a^2-a\xi_s .
\end{aligned}
\label{eq:Rs_def_VI}
\end{equation}
and
\begin{equation}
\begin{aligned}
\Theta_s(\theta;\xi_s,k_s)
&=
\frac{k_s}{B_s(\theta)}
-\frac{Q_s(\theta)^2}{\sin^2\theta},
\\
Q_s(\theta)&:=\xi_s-a\sin^2\theta .
\end{aligned}
\label{eq:Thetas_def_VI}
\end{equation}
These expressions do not simply import the Kerr formulas: they are the
separated potentials of the optical metrics truncated at first order. When
$\beta\to0$, $A_s,B_s\to1$, and both branches recover the usual null
structure.

The critical family is selected by imposing a double root of the radial
potential,
\begin{equation}
\mathcal R_s(r_c;\xi_s,k_s)=0,
\qquad
\partial_r\mathcal R_s(r_c;\xi_s,k_s)=0.
\label{eq:double_root_conditions_VI}
\end{equation}
This condition identifies the critical trajectories that organize the
projected shadow edge for each optical cone
\cite{GrenzebachPerlickLammerzahl2014,PerlickTsupko2022}.

To solve \eqref{eq:double_root_conditions_VI}, we write
\begin{equation}
\begin{aligned}
\Delta_c&:=\Delta(r_c),
&
A_{s,c}&:=A_s(r_c),\\[1mm]
\mathfrak D_s(r)&:=
\Delta'(r)A_s(r)-\Delta(r)A_s'(r) .
\end{aligned}
\label{eq:critical_shortcuts_VI}
\end{equation}
From the first condition in
\eqref{eq:double_root_conditions_VI},
\begin{equation}
k_s
=
\frac{A_{s,c}}{\Delta_c}\,P_{s,c}^2.
\label{eq:k_from_first_root_VI}
\end{equation}
The second condition uses
\begin{equation}
\left(\frac{\Delta}{A_s}\right)'
=
\frac{\Delta'A_s-\Delta A_s'}{A_s^2}
=
\frac{\mathfrak D_s}{A_s^2}.
\label{eq:Delta_over_A_derivative_VI}
\end{equation}
Since $P_s'(r)=2r$, the condition
$\partial_r\mathcal R_s=0$ gives
\begin{equation}
4r_cP_{s,c}
-
k_s
\frac{\mathfrak D_{s,c}}{A_{s,c}^2}
=
0,
\qquad
\mathfrak D_{s,c}:=\mathfrak D_s(r_c).
\label{eq:radial_derivative_intermediate_VI}
\end{equation}
Substituting \eqref{eq:k_from_first_root_VI} and excluding the degenerate
case $P_{s,c}=0$, one obtains
\begin{equation}
P_{s,c}
=
\frac{4r_c\Delta_c A_{s,c}}{\mathfrak D_{s,c}}.
\label{eq:Psc_solution_VI}
\end{equation}
Therefore, the critical constants are parametrized by the critical radius:
\begin{equation}
\xi_s(r_c)
=
\frac{
(r_c^2+a^2)\mathfrak D_{s,c}
-
4r_c\Delta_c A_{s,c}
}{
a\mathfrak D_{s,c}
},
\label{eq:xi_s_critical_VI}
\end{equation}
and
\begin{equation}
k_s(r_c)
=
\frac{
16r_c^2\Delta_c A_{s,c}^3
}{
\mathfrak D_{s,c}^2
}.
\label{eq:k_s_critical_VI}
\end{equation}
Each Fresnel root thus generates its own critical family.

The Maxwell limit provides the first check. When $\beta\to0$,
\begin{equation}
A_s\to1,
\qquad
\mathfrak D_s\to\Delta',
\label{eq:Maxwell_As_Ds_limit_VI}
\end{equation}
and both branches collapse to
\begin{equation}
\xi_0(r_c)
=
\frac{
(r_c^2+a^2)\Delta_c'
-
4r_c\Delta_c
}{
a\Delta_c'
},
\qquad
k_0(r_c)
=
\frac{16r_c^2\Delta_c}{(\Delta_c')^2}.
\label{eq:xi_k_0_limit_VI}
\end{equation}
This is the unique critical value of the background null cone.

The splitting can be read analytically by expanding only the optical
deformation,
\begin{equation}
A_s(r_c)=1+\beta\alpha_s r_c^2,
\qquad
A_s'(r_c)=2\beta\alpha_s r_c,
\label{eq:As_expansion_VI}
\end{equation}
while $\Delta(r)$ is kept as the exact background function. Then
\begin{equation}
\mathfrak D_{s,c}
=
\Delta_c'
+
\beta\alpha_s
\left(
r_c^2\Delta_c'
-
2r_c\Delta_c
\right)
+
\mathcal O(\beta^2).
\label{eq:Ds_expansion_VI}
\end{equation}
To show how this expansion enters $\xi_s$, we write
\begin{equation}
\mathfrak D_{s,c}=\Delta_c'+\beta\alpha_s d_c,
\qquad
d_c:=r_c^2\Delta_c'-2r_c\Delta_c .
\label{eq:dc_def_VI}
\end{equation}
Expanding \eqref{eq:xi_s_critical_VI} to first order gives
\begin{equation}
\begin{aligned}
\xi_s
={}&\xi_0
+\frac{\beta\alpha_s}{a}
\Biggl[
\frac{(r_c^2+a^2)d_c-4r_c^3\Delta_c}{\Delta_c'}
\\
&-\frac{\bigl[(r_c^2+a^2)\Delta_c'-4r_c\Delta_c\bigr]d_c}
{(\Delta_c')^2}
\Biggr]
+\mathcal O(\beta^2) .
\end{aligned}
\label{eq:xi_s_expansion_intermediate_VI}
\end{equation}
Using the definition of $d_c$, the terms in brackets reduce to
\begin{equation}
-\frac{8r_c^2\Delta_c^2}{(\Delta_c')^2},
\label{eq:xi_s_bracket_reduction_VI}
\end{equation}
and hence
\begin{equation}
\xi_s(r_c)
=
\xi_0(r_c)
-
\frac{
8\beta\alpha_s r_c^2\Delta_c^2
}{
a(\Delta_c')^2
}
+
\mathcal O(\beta^2).
\label{eq:xi_s_linear_VI}
\end{equation}
Since $\alpha_+-\alpha_-=12$, it follows that
\begin{equation}
\Delta\xi_{\rm br}(r_c)
:=
\xi_+(r_c)-\xi_-(r_c)
=
-\frac{
96\beta r_c^2\Delta_c^2
}{
a(\Delta_c')^2
}
+
\mathcal O(\beta^2).
\label{eq:Dxi_96_VI}
\end{equation}

For $k_s$, we use
\begin{equation}
A_{s,c}^3
=
1+3\beta\alpha_s r_c^2
+
\mathcal O(\beta^2),
\label{eq:Ac_cubed_expansion_VI}
\end{equation}
and
\begin{equation}
\mathfrak D_{s,c}^{-2}
=
(\Delta_c')^{-2}
\left[
1
-
2\beta\alpha_s
\frac{d_c}{\Delta_c'}
\right]
+
\mathcal O(\beta^2).
\label{eq:Ds_inverse_squared_expansion_VI}
\end{equation}
Substitution into \eqref{eq:k_s_critical_VI} gives
\begin{equation}
k_s
=
k_0
\left[
1+\beta\alpha_s
\left(
3r_c^2-2\frac{d_c}{\Delta_c'}
\right)
\right]
+
\mathcal O(\beta^2).
\label{eq:k_s_expansion_intermediate_VI}
\end{equation}
Finally, using $d_c=r_c^2\Delta_c'-2r_c\Delta_c$, one obtains
\begin{equation}
k_s(r_c)
=
k_0(r_c)
\left[
1+
\beta\alpha_s
\left(
r_c^2+\frac{4r_c\Delta_c}{\Delta_c'}
\right)
\right]
+
\mathcal O(\beta^2),
\label{eq:k_s_linear_VI}
\end{equation}
and consequently
\begin{equation}
\begin{aligned}
\Delta k_{\rm br}(r_c)
&:=k_+(r_c)-k_-(r_c)
\\
&=
12\beta k_0(r_c)
\left(
r_c^2+\frac{4r_c\Delta_c}{\Delta_c'}
\right)
+\mathcal O(\beta^2) .
\end{aligned}
\label{eq:Dk_12_VI}
\end{equation}
The points where $\mathfrak D_{s,c}=0$ or $\Delta_c'=0$ lie outside this
parametrization, because there the nondegenerate double root is no longer
uniformly described by
\eqref{eq:xi_s_critical_VI}--\eqref{eq:k_s_critical_VI}.

\subsection{Projection onto the local celestial sphere}
\label{subsec:finite_screen_projection}

The splittings $\Delta\xi_{\rm br}$ and $\Delta k_{\rm br}$ belong to the
space of critical constants. To convert them into an image diagnostic, each
family must be projected onto the celestial sphere of a local observer. We
consider an observer located at
\begin{equation}
O=(r_o,\theta_o),
\qquad
\Sigma_o:=r_o^2+a^2\cos^2\theta_o,
\qquad
\Delta_o:=\Delta(r_o),
\label{eq:observer_position_VI}
\end{equation}
with
\begin{equation}
\Delta_o>0,
\qquad
0<\theta_o<\pi .
\label{eq:observer_domain_VI}
\end{equation}
The projection is carried out in the orthonormal frame associated with the
principal tetrad of the background, as in finite-distance observer
constructions of black hole shadows
\cite{GrenzebachPerlickLammerzahl2014,PerlickTsupko2022}. This choice avoids
introducing impact parameters at infinity, which are not the natural
description for the García--Díaz branch considered here.

For each branch $s=\pm$, we evaluate the separated combinations at the
observation point,
\begin{equation}
\begin{aligned}
P_{o,s}(r_c)&:=r_o^2+a^2-a\xi_s(r_c),
\\
Q_{o,s}(r_c)&:=\xi_s(r_c)-a\sin^2\theta_o .
\end{aligned}
\label{eq:PoQo_def_VI}
\end{equation}
together with
\begin{equation}
\Theta_{o,s}(r_c)
:=
\frac{k_s(r_c)}{B_s(\theta_o)}
-
\frac{Q_{o,s}(r_c)^2}{\sin^2\theta_o},
\label{eq:Thetaos_def_VI}
\end{equation}
and
\begin{equation}
\mathcal R_{o,s}(r_c)
:=
P_{o,s}(r_c)^2
-
\frac{\Delta_o}{A_s(r_o)}\,k_s(r_c).
\label{eq:Ros_def_VI}
\end{equation}
The conditions
\begin{equation}
\Theta_{o,s}(r_c)\ge0,
\qquad
\mathcal R_{o,s}(r_c)\ge0
\label{eq:screen_reality_conditions_VI}
\end{equation}
select the critical radii whose projection reaches the observer with real
tetrad components.

After factoring out the energy $\mathcal E$, the tetrad covector of the
branch can be written as
\begin{equation}
p_{\hat0}^{(s)}
=
-\frac{\mathcal E\,P_{o,s}}
{\sqrt{\Delta_o\Sigma_o}},
\qquad
p_{\hat1}^{(s)}
=
\sigma_r\,
\frac{\mathcal E\,\sqrt{\mathcal R_{o,s}}}
{\sqrt{\Delta_o\Sigma_o}},
\label{eq:p01_screen_VI}
\end{equation}
\begin{equation}
p_{\hat2}^{(s)}
=
\varsigma\,
\frac{\mathcal E\,\sqrt{\Theta_{o,s}}}
{\sqrt{\Sigma_o}},
\qquad
p_{\hat3}^{(s)}
=
\frac{\mathcal E\,Q_{o,s}}
{\sqrt{\Sigma_o}\sin\theta_o}.
\label{eq:p23_screen_VI}
\end{equation}
Here $\sigma_r=\pm1$ fixes the radial orientation of the ray at the
observer, while $\varsigma=\pm1$ distinguishes the two halves of the
projected curve. In a comparison between branches, the same choice of
$\sigma_r$ and $\varsigma$ is taken for both optical cones.

In a birefringent theory, the observed direction is not obtained by directly
identifying the covector $p_\mu$ with the spatial direction of
propagation. The transport direction of each branch is given by the
Hamiltonian flow associated with the effective optical cone,
\begin{equation}
K_s^\mu
:=
\frac{dx^\mu}{d\lambda}
=
\frac{\partial\mathcal H_s}{\partial p_\mu}
=
\tilde g_s^{\mu\nu}p_\nu,
\label{eq:Ks_def_VI}
\end{equation}
in accordance with the geometrical description of propagation in NLED
\cite{NovelloEtAl2000,ObukhovRubilar2002}. Evaluated at the observer, we
define
\begin{equation}
\Upsilon_{o,s}:=\Upsilon_s(r_o,\theta_o),
\qquad
\Phi_{o,s}:=\Phi_s(r_o,\theta_o),
\label{eq:UpsilonPhi_observer_def_VI}
\end{equation}
so that, in the optical tetrad used here,
\begin{equation}
K_s^{\hat0}
=
-\Upsilon_{o,s}\,p_{\hat0}^{(s)},
\qquad
K_s^{\hat1}
=
\Upsilon_{o,s}\,p_{\hat1}^{(s)},
\label{eq:K01_screen_VI}
\end{equation}
\begin{equation}
K_s^{\hat2}
=
\Phi_{o,s}\,p_{\hat2}^{(s)},
\qquad
K_s^{\hat3}
=
\Phi_{o,s}\,p_{\hat3}^{(s)}.
\label{eq:K23_screen_VI}
\end{equation}

The observed direction is defined as the normalized spatial part of
$K_s^{\hat a}$,
\begin{equation}
\mathbf n_s(r_c,\varsigma)
:=
\frac{
\bigl(
K_s^{\hat1},
K_s^{\hat2},
K_s^{\hat3}
\bigr)
}{
\sqrt{
(K_s^{\hat1})^2+
(K_s^{\hat2})^2+
(K_s^{\hat3})^2
}
}.
\label{eq:ns_def_VI}
\end{equation}
Thus, each admissible critical radius and each choice of $\varsigma$
determine a point on the local celestial sphere of the observer.

To represent the curves, we use the stereographic projection
\begin{equation}
X_s(r_c,\varsigma)
=
\frac{2\,n_s^{\hat3}(r_c,\varsigma)}
{1+n_s^{\hat1}(r_c,\varsigma)},
\qquad
Y_s(r_c,\varsigma)
=
\frac{2\,n_s^{\hat2}(r_c,\varsigma)}
{1+n_s^{\hat1}(r_c,\varsigma)}.
\label{eq:XYs_stereo_VI}
\end{equation}
The visible interval of the branch $s$ is defined by
\begin{equation}
\begin{aligned}
\mathcal I_s
:=\Bigl\{\,r_c>r_+\; ;\;&
\mathfrak D_{s,c}\neq0,
\quad
\Theta_{o,s}\ge0,
\quad
\mathcal R_{o,s}\ge0,
\\
&\Upsilon_{o,s}>0,
\quad
\Phi_{o,s}>0
\Bigr\} .
\end{aligned}
\label{eq:Is_visible_VI}
\end{equation}
where $r_+$ denotes the outer horizon, when it exists. The condition
$\mathfrak D_{s,c}\neq0$ excludes degenerate points of the critical
parametrization, while the remaining inequalities guarantee the reality of
the projection and the Lorentzian signature of the optical representative.

\begin{figure*}[tbp]
\centering
\includegraphics[width=0.92\textwidth]{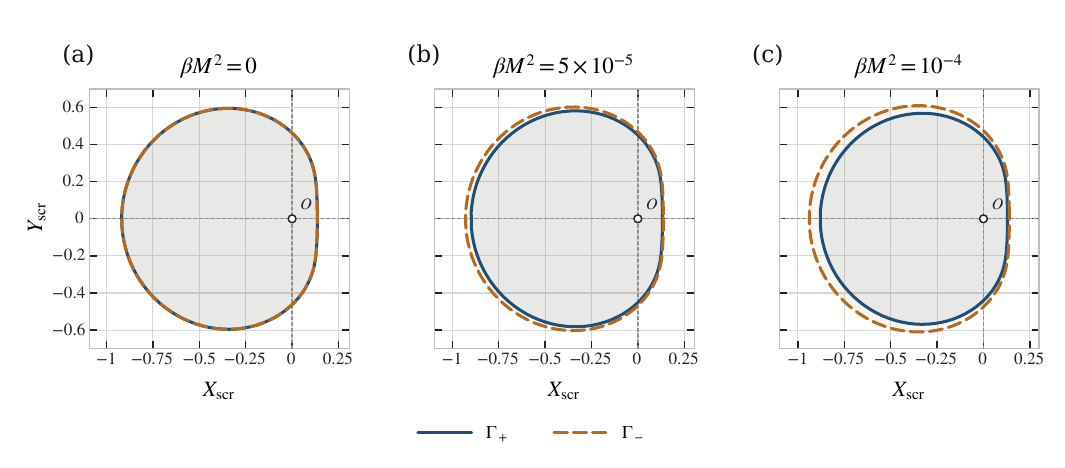}
\caption{
Critical contours $\Gamma_+$ and $\Gamma_-$ projected onto the local
screen for a scan in the nonlinear coupling. We fix
$M=1$, $p=0.2$, $q=0.3$, $r_o=8$,
$\theta_o=85^\circ$, and $a/M=0.93$. The solid curve corresponds to
$\Gamma_+$, and the dashed curve to $\Gamma_-$. Panel (a) shows the
Maxwell limit, where the two Fresnel roots collapse onto the same optical
cone and the contours coincide. In panels (b) and (c), increasing
$\beta M^2$ breaks this degeneracy and separates the two critical
families. The opening between the curves is the direct geometrical imprint
of constitutive birefringence.
}
\label{fig:gamma_beta_scan}
\end{figure*}

\begin{figure*}[tbp]
\centering
\includegraphics[width=0.92\textwidth]{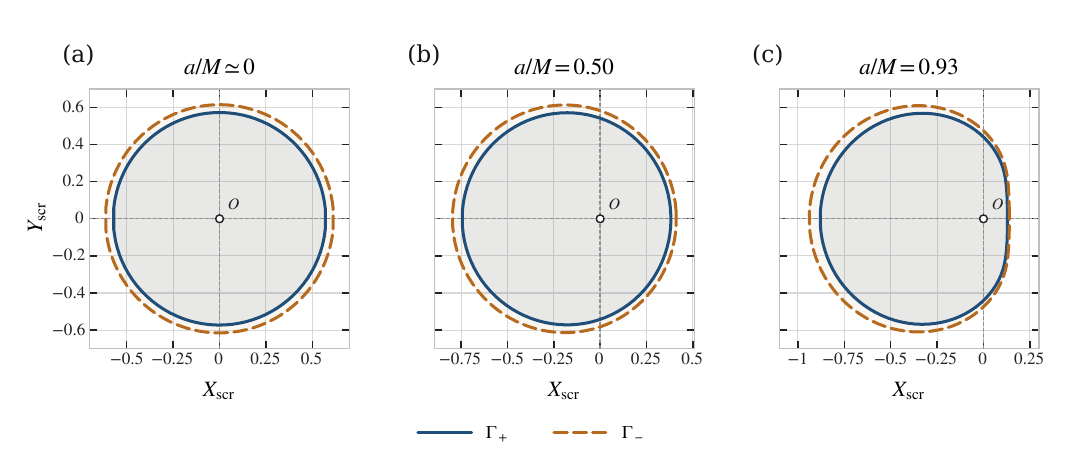}
\caption{
Critical contours $\Gamma_+$ and $\Gamma_-$ projected onto the local
screen for a scan in the rotation parameter. We fix
$M=1$, $p=0.2$, $q=0.3$, $r_o=8$,
$\theta_o=85^\circ$, and $\beta M^2=10^{-4}$. The solid curve
corresponds to $\Gamma_+$, and the dashed curve to $\Gamma_-$. Panel (a)
shows the near-nonrotating case, computed with the small spin regulator
$a/M=10^{-3}$; at the plotting resolution, the splitting is
indistinguishable from the spherical limit and appears as an almost
concentric separation between branches. As $a/M$ increases, panels (b)
and (c), the axial geometry of the background shifts the visible critical
family and redistributes the splitting across the screen. Rotation does
not generate birefringence, but it controls how the constitutive
separation between optical cones appears in the local projection.
}
\label{fig:gamma_spin_scan}
\end{figure*}

With these quantities, we formally define the critical contour of the
optical branch $s$ on the local celestial sphere as
\begin{equation}
\Gamma_s^{\rm cel}
:=
\left\{
\mathbf n_s(r_c,\varsigma)\in S^2_o
\; ; \;
r_c\in\mathcal I_s,\;
\varsigma=\pm1
\right\},
\qquad
s=\pm ,
\label{eq:Gamma_cel_def_VI}
\end{equation}
and its image on the observer's stereographic screen as
\begin{equation}
\Gamma_s
:=
\left\{
\bigl(X_s(r_c,\varsigma),Y_s(r_c,\varsigma)\bigr)
\; ; \;
r_c\in\mathcal I_s,\;
\varsigma=\pm1
\right\}.
\label{eq:Gamma_screen_def_VI}
\end{equation}
In what follows, $\Gamma_s$ will denote the projected contour on the local
screen; we will write $\Gamma_s^{\rm cel}$ only when it is necessary to
emphasize the curve on the celestial sphere.

The birefringent comparison is performed on
\begin{equation}
\mathcal I_{\rm br}:=\mathcal I_+\cap\mathcal I_-,
\label{eq:Ibr_visible_VI}
\end{equation}
where both branches are simultaneously defined. The angular separation
between the two branches is measured by
\begin{equation}
\Delta_{\rm cel}(r_c,\varsigma)
:=
\arccos
\left[
\mathbf n_+(r_c,\varsigma)\cdot\mathbf n_-(r_c,\varsigma)
\right],
\qquad
r_c\in\mathcal I_{\rm br}.
\label{eq:Delta_cel_def_VI}
\end{equation}
This quantity is intrinsic to the local celestial sphere and does not depend
on the two-dimensional chart used to draw the image. On the stereographic
screen, the projected separation between branches is
\begin{equation}
\begin{aligned}
\Delta_{\rm scr}(r_c,\varsigma)
:={}&
\Bigl\{
\bigl[X_+(r_c,\varsigma)-X_-(r_c,\varsigma)\bigr]^2
\\
&+
\bigl[Y_+(r_c,\varsigma)-Y_-(r_c,\varsigma)\bigr]^2
\Bigr\}^{1/2} .
\end{aligned}
\label{eq:Delta_scr_def_VI}
\end{equation}
While $\Delta_{\rm cel}$ measures the intrinsic angular splitting,
$\Delta_{\rm scr}$ describes the separation in the stereographic chart
used to visualize the contours.

On this domain we define
\begin{equation}
\begin{aligned}
\Delta_{\rm cel}^{\rm max}
&:=
\max_{\substack{r_c\in\mathcal I_{\rm br}\\ \varsigma=\pm1}}
\Delta_{\rm cel}(r_c,\varsigma),\\[1mm]
\Delta_{\rm scr}^{\rm max}
&:=
\max_{\substack{r_c\in\mathcal I_{\rm br}\\ \varsigma=\pm1}}
\Delta_{\rm scr}(r_c,\varsigma) .
\end{aligned}
\label{eq:Delta_cel_scr_max_VI}
\end{equation}
The maximum $\Delta_{\rm cel}^{\rm max}$ is intrinsic to the celestial
sphere. By contrast, $\Delta_{\rm scr}^{\rm max}$ depends on the chosen
screen chart and will be used only as a visualization diagnostic. In the
Maxwell limit, the two optical roots collapse to the same cone, the critical
families coincide,
\begin{equation}
\Gamma_+\to\Gamma_-,
\qquad
\Gamma_+^{\rm cel}\to\Gamma_-^{\rm cel},
\label{eq:Gamma_Maxwell_limit_VI}
\end{equation}
and both splittings vanish.

The previous construction translates the bifurcation of the optical cones into
two critical contours on the observer's local screen. In the linear limit the
two branches share the same critical family; away from that limit, the
constitutive response separates the optical potentials and produces
$\Gamma_+\neq\Gamma_-$. Figs.~\ref{fig:gamma_beta_scan} and
\ref{fig:gamma_spin_scan} display this separation before it is reduced to
global observables. The first figure isolates the role of the nonlinear
coupling, whereas the second shows how rotation reorganizes the same
geometrical signal.

\subsection{Effective diameter and birefringent geometric width}
\label{subsec:birefringent_diameter_width}

The separation between $\Gamma_+$ and $\Gamma_-$ is not merely a local
difference between celestial directions: it affects the full morphology of
the projected contour. The maximum $\Delta_{\rm cel}^{\rm max}$ captures
the largest angular separation between branches, but it does not by itself
describe the change in global scale or the transverse distribution of the
separation. To characterize this global information, we now introduce two
screen-based quantities: a relative diameter splitting and an effective
geometric width. In horizon-scale image analyses, quantities such as the
diameter and the fractional width are used to characterize the observed
structure \cite{EHT2019VI,EHT2022SgrAIV}. Here we use their purely
geometrical analogues for the critical contours.

We work on the common visibility interval $\mathcal I_{\rm br}$, defined in
\eqref{eq:Ibr_visible_VI}. For each optical branch $s=\pm$, we introduce
the screen vector
\begin{equation}
\begin{aligned}
\mathbf X_s(r_c,\varsigma)
&:=
\bigl(X_s(r_c,\varsigma),Y_s(r_c,\varsigma)\bigr),
\\
r_c&\in\mathcal I_{\rm br},
\qquad
\varsigma=\pm1 .
\end{aligned}
\label{eq:Xvector_s_def_VI}
\end{equation}
Scanning $\mathcal I_{\rm br}$, together with the two choices of
$\varsigma$, gives the projected contour $\Gamma_s$. We assume that the
common domain generates closed and regular curves on the screen. If, for a
given choice of parameters, only a visible arc is obtained, the same
definitions should be understood as restricted to that arc.

The line element on the local screen is defined as
\begin{equation}
d\ell_s
:=
\sqrt{dX_s^2+dY_s^2},
\qquad
L_s:=\oint_{\Gamma_s}d\ell_s .
\label{eq:dells_Ls_def_VI}
\end{equation}
With this measure, the geometrical center of the contour is taken as the
arc-length average,
\begin{equation}
\mathbf C_s
:=
\frac{1}{L_s}
\oint_{\Gamma_s}
\mathbf X_s\,d\ell_s .
\label{eq:Cs_def_VI}
\end{equation}
The projected mean radius of the branch $s$ is defined by
\begin{equation}
\bar R_s
:=
\frac{1}{L_s}
\oint_{\Gamma_s}
\left|
\mathbf X_s-\mathbf C_s
\right|\,d\ell_s ,
\label{eq:Rbar_s_def_VI}
\end{equation}
and the associated effective diameter by
\begin{equation}
d_s:=2\bar R_s .
\label{eq:ds_def_VI}
\end{equation}
This definition does not assume circularity. For a circular curve it
reproduces the usual diameter; for a deformed contour it gives an average
size scale on the screen. Thus, $d_s$ should be understood as a geometrical
measure of the critical contour, not as an image diameter already processed
by emission, radiative transfer, or instrumental response.

The first global birefringence observable is the relative diameter splitting,
\begin{equation}
\delta d_{\rm br}
:=
\frac{2|d_+-d_-|}{d_++d_-}.
\label{eq:delta_d_br_def_VI}
\end{equation}
This quantity measures whether the two optical branches, in addition to
separating pointwise on the celestial sphere, predict different global
scales. In the Maxwell limit,
\begin{equation}
d_+\to d_-,
\qquad
\delta d_{\rm br}\to0
\qquad
(\beta\to0).
\label{eq:delta_d_br_Maxwell_VI}
\end{equation}

If the contours $\Gamma_+$ and $\Gamma_-$ are too close to be resolved as
two separate curves, their separation can be described as a thin band around
a mean contour. We define that contour by
\begin{equation}
\bar{\mathbf X}(r_c,\varsigma)
:=
\frac12
\left[
\mathbf X_+(r_c,\varsigma)+\mathbf X_-(r_c,\varsigma)
\right],
\label{eq:Xbar_def_VI}
\end{equation}
and denote by $\bar\Gamma$ the curve generated by $\bar{\mathbf X}$. The
critical curve is a geometrical object, whereas an observed image depends on
the emitting plasma, radiative transfer, and instrumental response; hence
the relation between critical curves and observed appearance is not direct
\cite{GrallaHolzWald2019,EHT2019V}. Here we define only the geometrical
contribution associated with the separation between the two critical
contours.

With line element
\begin{equation}
d\bar\ell
:=
\left|
d\bar{\mathbf X}
\right|,
\qquad
\bar L:=\oint_{\bar\Gamma}d\bar\ell ,
\label{eq:dellbar_Lbar_def_VI}
\end{equation}
at regular points of the mean contour we define the unit tangent vector
\begin{equation}
\hat{\mathbf t}
:=
\frac{d\bar{\mathbf X}}
{\left|d\bar{\mathbf X}\right|},
\label{eq:tangent_mean_def_VI}
\end{equation}
and the local oriented normal on the screen,
\begin{equation}
\hat{\mathbf m}
:=
\bigl(-\hat t_Y,\hat t_X\bigr).
\label{eq:normal_mean_def_VI}
\end{equation}
The separation between the branches is written as
\begin{equation}
\delta\mathbf X_{\rm br}(r_c,\varsigma)
:=
\mathbf X_+(r_c,\varsigma)-\mathbf X_-(r_c,\varsigma),
\label{eq:deltaX_vector_br_def_VI}
\end{equation}
and its normal component with respect to the mean contour is
\begin{equation}
\delta X_\perp(r_c,\varsigma)
:=
\delta\mathbf X_{\rm br}(r_c,\varsigma)\cdot\hat{\mathbf m}(r_c,\varsigma).
\label{eq:deltaX_perp_def_VI}
\end{equation}
This component is the one that contributes to the local geometrical
thickness. The tangential component, by contrast, shifts points along the
contour and does not by itself produce transverse broadening.

We then define the effective birefringent geometric width by
\begin{equation}
w_{\rm br}^2
:=
\frac{1}{4\bar L}
\oint_{\bar\Gamma}
\bigl[\delta X_\perp\bigr]^2
\,d\bar\ell .
\label{eq:wbr_def_VI}
\end{equation}
The factor $1/4$ appears because $\delta X_\perp$ measures the total
normal separation between the two branches.  If the branches are distributed
approximately symmetrically around the mean contour, each branch lies at a
distance $|\delta X_\perp|/2$ from $\bar\Gamma$.  Thus,
$w_{\rm br}$ measures the root-mean-square (rms) half-width induced by
birefringence, while $2w_{\rm br}$ represents the full rms normal
separation between the two contours.

To obtain a dimensionless quantity, we normalize this width by the mean
diameter
\begin{equation}
d_0
:=
\frac12(d_++d_-),
\label{eq:d0_mean_def_VI}
\end{equation}
and define
\begin{equation}
f_{\rm br}
:=
\frac{w_{\rm br}}{d_0}.
\label{eq:fbr_def_VI}
\end{equation}
This diagnostic is the geometrical analogue of a fractional width. If the two
critical curves are not resolved separately, $f_{\rm br}$ quantifies the
minimum contribution that the separation of the optical cones can make to
the effective width of the projected structure. In the linear limit,
\begin{equation}
\begin{aligned}
\mathbf X_+(r_c,\varsigma)&\to\mathbf X_-(r_c,\varsigma),
\\
w_{\rm br}&\to0,
\qquad
f_{\rm br}\to0,
\qquad
(\beta\to0) .
\end{aligned}
\label{eq:fbr_Maxwell_limit_VI}
\end{equation}

The construction leaves three main geometrical diagnostics:
\begin{equation}
\Delta_{\rm cel}^{\rm max},
\qquad
\delta d_{\rm br},
\qquad
f_{\rm br}.
\label{eq:main_geometric_observables_summary_VI}
\end{equation}
The first one measures the maximum local angular splitting and is defined
directly on the observer's celestial sphere.  The other two condense the
same separation into global screen scales: $\delta d_{\rm br}$ measures
whether the two branches predict different effective diameters, whereas
$f_{\rm br}$ measures the geometrical width associated with an unresolved
splitting.  By contrast, $\Delta_{\rm scr}^{\rm max}$ is kept only as a
diagnostic of the stereographic chart used to visualize the contours.

To complement the maximum angular separation, which may be sensitive to the
endpoints of the visible interval, we also report the discrete average
\begin{equation}
\langle\Delta_{\rm cel}\rangle
:=
\frac{1}{N_{\rm br}}
\sum_{(r_c,\varsigma)\in\mathcal I_{\rm br}^{\rm num}}
\Delta_{\rm cel}(r_c,\varsigma).
\label{eq:Delta_cel_mean_def_VI}
\end{equation}
Here $\mathcal I_{\rm br}^{\rm num}$ denotes the set of numerical points
in the common domain used to trace both branches simultaneously, including
the two choices of $\varsigma$, and $N_{\rm br}$ is its cardinality.
This average does not replace $\Delta_{\rm cel}^{\rm max}$; it only
summarizes the typical size of the angular splitting along the common
contour.

To assess the geometrical stability of these diagnostics, we perform three
representative scans within the perturbative domain
$\epsilon_{\rm pert}\leq0.05$.  In all of them we fix
\[
M=1,\qquad p=0.2,\qquad q=0.3,\qquad r_o=8,
\]
and vary, respectively, the nonlinear coupling $\beta$, the rotation
parameter $a/M$, and the observer inclination $\theta_o$.  The complete
numerical values, together with the rationale for the reference dyonic
sector, are collected in Appendix~\ref{app:numerical_scans}.

The qualitative reading of these scans is direct.  In the Maxwell limit, the
contours $\Gamma_+$ and $\Gamma_-$ collapse into a single critical curve,
and the diagnostics $\Delta_{\rm cel}^{\rm max}$, $\delta d_{\rm br}$,
and $f_{\rm br}$ vanish simultaneously.  When the nonlinear constitutive
response is turned on, the two optical branches generate distinct critical
families: $\beta$ controls the opening between the cones, rotation
redistributes the signal over the screen, and inclination changes its local
projective realization without removing the splitting.

This completes the geometrical chain constructed in this section.  The local
constitutive response separates the Fresnel cones; each cone defines an
effective optical metric; Carter-type separability allows one to construct
branch-dependent critical families; and the local projection maps those
families into two screen contours.  The contours $\Gamma_\pm$ are therefore
critical curves of the effective optical geometries.  They are not luminous
rings and do not replace an accretion image; rather, they set the geometrical
support on which emission, plasma effects, and radiative transfer can be
incorporated in a subsequent step.


\section{Physical discussion and scope}
\label{sec:discussion}

The calculation has a simple physical meaning.  The rotating
Garc\'ia--D\'iaz metric gives the gravitational background, but in NLED the
propagation of electromagnetic perturbations is not fixed by this metric
alone.  The perturbation also feels the local constitutive law of the
nonlinear electromagnetic field.  For this reason the pair $(F,P)$ is not only
a technical way to write the solution.  It contains the response that fixes
the optical cones \cite{Plebanski1970,SalazarGarciaPlebanski1987,
SalazarGarciaPlebanski1989,Boillat1970,NovelloEtAl2000,ObukhovRubilar2002}.

This point is especially useful for the Garc\'ia--D\'iaz family.  The same
exact branch contains the metric, the mixed potentials and the constitutive
structure of the Einstein--NLED system \cite{GarciaDiaz2021,
GarciaDiaz2022Adsds,AyonBeato2024}.  Therefore, the optical metrics used in
this work are not phenomenological deformations of Kerr or Kerr--Newman.  They
are obtained from the local response matrix reconstructed from the exact
fields.  In this sense, the birefringent splitting belongs to the solution
itself.

At the perturbative order considered here, the Fresnel quartic factorizes into
two quadratic branches.  In the Maxwell limit the two branches collapse to the
same null cone.  When the nonlinear response is active, they define two
different effective optical metrics.  In the principal tetrad, the deformation
acts differently on the temporal--radial and angular blocks, and the tensor
$F^\mu{}_{\alpha}F^{\alpha\nu}$ gives the algebraic direction of the effect.
This is the local form of constitutive birefringence.

The next important point is separability.  The two optical metrics are not the
background metric, but at first order they still keep enough Kerr-type
structure to separate the Hamilton--Jacobi equation.  Thus each branch has its
own radial potential, angular potential, separation constant and critical
family.  Birefringence therefore does not appear here as a loss of the
critical problem.  It appears as a doubling of it.

This doubling becomes visible only after projection.  In a rotating geometry,
the shadow edge is not controlled by one circular orbit.  It is controlled by
a family of critical orbits and by the way this family is projected on the
screen of the observer \cite{Carter1968HJ,Chandrasekhar1983,
GrenzebachPerlickLammerzahl2014,PerlickTsupko2022}.  Each optical branch
produces its own projected contour, $\Gamma_+$ or $\Gamma_-$.  The separation
between these contours is the geometrical imprint of birefringence.  We
measure it by the maximum angular separation, the relative diameter shift and
the normalized geometrical width.  These are not flux observables.  They are
clean geometrical quantities defined before choosing any accretion model.

The Maxwell limit gives a useful check.  When $\beta\to0$, the two Fresnel
roots become the same, and the diagnostics $\Delta_{\rm cel}^{\rm max}$,
$\delta d_{\rm br}$ and $f_{\rm br}$ all go to zero.  Thus the splitting is
not produced by the observer tetrad, by the screen coordinates or by a choice
of parametrization.  It comes from the nonlinear constitutive response.

The use of a finite-distance observer is also part of the construction.  For
$\beta\neq0$, the branch is not asymptotically flat in the usual sense, so it
is not natural to define the shadow only through impact parameters at
infinity.  A local orthonormal tetrad gives a well-defined projection of the
critical families.  This places the construction in the same spirit as
finite-distance treatments of shadows and critical curves in rotating
geometries \cite{GrenzebachPerlickLammerzahl2014,PerlickTsupko2022,
CunhaHerdeiroRadu2017,GrallaLupsascaMarrone2020,HimwichEtAl2020,
PaugnatEtAl2022}.

The numerical scans support this interpretation.  In the controlled regime
$\epsilon_{\rm pert}\leq0.05$, the splitting grows with the nonlinear
coupling.  Rotation does not create birefringence by itself, but it
redistributes the separation along the projected critical family.  Changing
the observer inclination modifies the projection, but it does not remove the
separation between $\Gamma_+$ and $\Gamma_-$.  This is why the signal is best
understood as constitutive in origin, redistributed by rotation and then seen
through the local observer projection.

There is also a useful connection with other NLED observables.  The recent
study of quasinormal modes in a static Pleba\'nski-type NLED black hole showed
that the nonlinear electromagnetic sector can affect the dynamical
perturbation problem \cite{Fathi:2025jrk}.  Here we find the complementary
result in geometrical optics: the same type of constitutive structure can also
affect the optical cones and the critical contours.  These two directions are
different, but they share the same physical lesson.  In Einstein--NLED
systems, the electromagnetic response is not a passive background detail.

The present work is still a geometrical study.  The curves $\Gamma_+$ and
$\Gamma_-$ are critical contours of two effective optical geometries.  They
are not yet full images produced by an accretion flow.  A complete image would
require emission, absorption, polarized radiative transfer, plasma effects and
instrumental response \cite{EHT2019I,EHT2019V,EHT2019VI,EHT2022SgrAI,
EHT2022SgrAIV,EHT2022SgrAV,EHT2022SgrAVI,MoscibrodzkaFalcke2013,
DexterAgol2009,VincentEtAl2011,YounsiEtAl2012,PsaltisEtAl2020}.  The role of
this paper is to isolate the geometrical support on which such a birefringent
image would have to be built.

The final message is therefore direct.  NLED does not only change the
stress-energy tensor that sources the geometry.  In the geometrical-optics
limit, it also changes the characteristic structure followed by the
electromagnetic perturbation.  In the rotating Garc\'ia--D\'iaz black hole,
this change can be followed explicitly from the local constitutive response to
the Fresnel cones, the separated optical dynamics and the projected critical
contours on the observer screen.


\section{Summary and conclusions}
\label{sec:conclusions}

We have studied high-frequency electromagnetic propagation in the rotating
Garc\'ia--D\'iaz black hole of Einstein gravity coupled to NLED.  The aim was
to follow one complete chain: from the exact mixed potentials of the solution,
to the local constitutive response, to the Fresnel optical metrics, and finally
to the critical contours seen by a finite-distance observer.

The first step was the reconstruction of the constitutive branch.  In the
principal tetrad, the electromagnetic sector is described by four aligned
scalars, $E$, $B$, $D$ and $H$.  In the Maxwell limit one has $D=E$ and
$H=B$.  For nonzero nonlinear coupling this equality is deformed by the
factors $\kappa_r$ and $\kappa_\theta$.  This allowed us to write the regular
branch as a local map $(D,B)\mapsto(E,H)$ and to obtain the response matrix.
The optical geometry was then derived from this response; it was not assumed
independently.

The second step was the Fresnel construction.  At the perturbative order used
here, the Fresnel quartic splits into two quadratic branches.  In the Maxwell
limit these branches become the same light cone.  Away from that limit, they
define two effective optical metrics.  The deformation is controlled by the
quadratic tensor of the background electromagnetic field, and it distinguishes
the temporal--radial block from the angular block in the principal tetrad.

The third step was the separability analysis.  Although the optical metrics
differ from the spacetime metric, both branches admit a Carter-type separation
at first order.  For each branch $s=\pm$, one can define a radial potential,
an angular potential, critical constants and a family of unstable critical
orbits.  Thus the nonlinear response does not remove the integrable structure
at this order.  Instead, it produces two branch-dependent copies of the
critical problem.

The fourth step was the local projection.  Since the branch is not
asymptotically flat for $\beta\neq0$, we projected the critical families on
the celestial sphere of a finite-distance observer.  This gives two contours,
$\Gamma_+$ and $\Gamma_-$.  They coincide when $\beta=0$ and separate when the
NLED response is active.  We quantified the separation by the maximum angular
distance between the contours, the relative diameter shift and the normalized
birefringent width.  These quantities isolate the geometrical part of the
effect before any emission model is introduced.

The numerical scans give the same picture in a more concrete way.  Increasing
$\beta$ opens the two branches.  Rotation redistributes the separation on the
screen, without being the source of birefringence by itself.  Changing the
observer inclination changes the projection, but the two-contour structure
remains.  In the Maxwell limit all the diagnostics vanish at the same time.
This confirms that the splitting is not a coordinate or projection artifact;
it is produced by the constitutive response of the nonlinear electromagnetic
sector.

The main conclusion is that the rotating Garc\'ia--D\'iaz solution gives a
clear example where the internal electromagnetic response of an Einstein--NLED
black hole reaches the observable critical geometry.  A local property of the
field--excitation relation becomes a splitting of the optical cones, then a
splitting of the critical families, and finally a separation between two
projected contours on the observer screen.

Several extensions are natural.  The first one is to include emission and
polarized radiative transfer on the two optical branches, in order to see how
$\Gamma_+$ and $\Gamma_-$ would be illuminated in synthetic images.  A second
one is to go beyond the perturbative optical reconstruction and test the
birefringent structure with direct numerical integration of the effective
characteristics.  A third one is to compare this rotating branch with static
and other Einstein--NLED solutions, so that one can separate the effects of
rotation, constitutive anisotropy and strong gravity.

Overall, the result shows that a rotating Einstein--NLED solution can keep
enough separable structure to allow analytic control, while at the same time
producing a genuine birefringent splitting of the critical contours.  This is
the central point of the paper: the local constitutive response can be carried,
in a controlled way, all the way to geometrical diagnostics on the observer
screen.

\section*{Code and data availability}
The computational material associated with this work is available in the
companion repository~\cite{GuzmanGDBirefringenceCode}. It contains the
notebook and numerical outputs used to reproduce the figures, tables and
parameter scans reported in Appendix~\ref{app:numerical_scans}.

\begin{acknowledgments}
M.F. acknowledges financial support from
Agencia Nacional de Investigación y Desarrollo (ANID)
through the FONDECYT postdoctoral Grant No.
3260029.  J.R.V. is partially supported by Centro de F\'isica Teórica de Valparaíso (CeFiTeV). 
\end{acknowledgments}

\appendix

\section{Numerical control of the birefringent splitting}
\label{app:numerical_scans}

This appendix presents the numerical control of the geometrical diagnostics
introduced in Sec.~\ref{sec:birefringent_critical_curves}.  Its purpose is
to document how the splitting between the critical families
$\Gamma_+$ and $\Gamma_-$ behaves under representative variations of the
nonlinear coupling, the rotation, and the observer inclination.  In this
way, the tables complement the analytical construction of the main text
without interrupting its development.

In all scans we fix
\begin{equation}
M=1,
\qquad
p=0.2,
\qquad
q=0.3,
\qquad
r_o=8,
\label{eq:app_fixed_parameters}
\end{equation}
and retain only configurations satisfying
\begin{equation}
\epsilon_{\rm pert}\leq0.05 .
\label{eq:app_epsilon_bound}
\end{equation}
This restriction keeps the calculation within the domain where the
perturbative approximation to the optical metrics remains controlled.

The choice of a fixed dyonic sector is deliberate.  In this branch, the
charges enter the radial function through $Q_c^2=p^2+q^2$ and also set the
constitutive scale appearing in the optical cones.  The linear fields
$E_0$ and $B_0$ are homogeneous of degree one in $(p,q)$, whereas the
perturbative optical coefficients contain the factor $1/Q_c^2$.  Therefore,
under a common rescaling $(p,q)\to\lambda(p,q)$, the products
$\nu_sE_0^2$ and $\nu_sB_0^2$, which control the local deformation of the
optical metrics, remain invariant at fixed geometry.  The remaining effect of
varying the total charge scale is mainly channeled through $\Delta(r)$, the
horizon position, the perturbative domain, and the visible portion of the
critical families.  For this reason, the tables fix a representative dyonic
sector and separate three effects more directly: the scan in $\beta$
isolates the constitutive opening, the scan in $a/M$ displays the
rotational redistribution, and the scan in $\theta_o$ measures the
dependence on the local projection.

In the tables, the maximum angular separation
$\Delta_{\rm cel}^{\rm max}$ and the discrete average
$\langle\Delta_{\rm cel}\rangle$ are shown multiplied by $10^4$.  The
relative diameter shift $\delta d_{\rm br}$ and the normalized width
$f_{\rm br}$ are shown multiplied by $10^2$.  The quantity
$\Delta_{\rm scr}^{\rm max}$ is kept as a diagnostic of the stereographic
chart used to visualize the contours, while $d_0$ and $w_{\rm br}$ are
reported in the geometrical units of the local screen.

\subsection{Constitutive opening}
\label{app:beta_scan}

The first scan varies the nonlinear coupling while keeping
$a/M=0.93$ and $\theta_o=85^\circ$ fixed.  This scan isolates the role of
the constitutive response.  In the Maxwell limit the two branches collapse
into a single contour, whereas for $\beta\neq0$ a geometrical separation
between $\Gamma_+$ and $\Gamma_-$ appears.  Table~\ref{tab:app_beta_scan_geometric}
shows this behavior directly: all birefringent diagnostics vanish for
$\beta=0$, and then grow as the nonlinear coupling is increased.

\begin{widetext}
\begingroup
\renewcommand{\arraystretch}{1.04}
\setlength{\tabcolsep}{3.8pt}

\begin{center}
\refstepcounter{table}\label{tab:app_beta_scan_geometric}
\begin{minipage}{0.96\textwidth}
\footnotesize
\textbf{TABLE~\thetable.} Scan in the nonlinear coupling.  We fix $a/M=0.93$ and
$\theta_o=85^\circ$.  The first row is the Maxwell limit.  All rows satisfy
$\epsilon_{\rm pert}\leq0.05$.
\end{minipage}
\vspace{2pt}

\footnotesize
\begin{ruledtabular}
\begin{tabular*}{0.96\textwidth}{@{\extracolsep{\fill}}cccccccc}
$10^4\beta M^2$ &
$10^4\Delta_{\rm cel}^{\rm max}$ &
$10^4\langle\Delta_{\rm cel}\rangle$ &
$\Delta_{\rm scr}^{\rm max}$ &
$d_0$ &
$10^2\delta d_{\rm br}$ &
$w_{\rm br}$ &
$10^2f_{\rm br}$
\\
\hline
0.00 & 0.000    & 0.000   & 0.00000 & 1.13823 & 0.0000 & 0.00000 & 0.0000 \\
0.30 & 601.123  & 109.375 & 0.07267 & 1.13439 & 2.0639 & 0.00622 & 0.5486 \\
0.50 & 820.870  & 180.936 & 0.09912 & 1.13195 & 3.4393 & 0.01026 & 0.9067 \\
0.70 & 996.341  & 251.676 & 0.12017 & 1.12960 & 4.8118 & 0.01423 & 1.2600 \\
1.00 & 1212.097 & 356.445 & 0.14595 & 1.12625 & 6.8701 & 0.02008 & 1.7825 \\
\end{tabular*}
\end{ruledtabular}
\end{center}

\end{widetext}

\subsection{Rotational redistribution}
\label{app:spin_scan}

The second scan varies $a/M$ while keeping $\beta M^2=10^{-4}$ and
$\theta_o=85^\circ$ fixed.  This scan does not measure the generation of
birefringence, which is already present for $\beta\neq0$, but rather how
rotation redistributes the signal on the local screen.  As shown in
Table~\ref{tab:app_spin_scan_geometric}, the constitutive response already
separates the branches in the nonrotating limit.  When the spin is changed,
the visible critical family changes its shape and location, so the maximum
separation does not need to be monotonic, while the global scale of the
splitting remains comparable.

\begin{widetext}
    \begin{center}
\refstepcounter{table}\label{tab:app_spin_scan_geometric}
\begin{minipage}{0.96\textwidth}
\footnotesize
\textbf{TABLE~\thetable.} Scan in the rotation parameter.  We fix
$\beta M^2=10^{-4}$ and $\theta_o=85^\circ$.  The row $a/M=0$ is evaluated
with the spherical branch.  All rows satisfy $\epsilon_{\rm pert}\leq0.05$.
\end{minipage}
\vspace{2pt}

\footnotesize
\begin{ruledtabular}
\begin{tabular*}{0.96\textwidth}{@{\extracolsep{\fill}}cccccccc}
$a/M$ &
$10^4\Delta_{\rm cel}^{\rm max}$ &
$10^4\langle\Delta_{\rm cel}\rangle$ &
$\Delta_{\rm scr}^{\rm max}$ &
$d_0$ &
$10^2\delta d_{\rm br}$ &
$w_{\rm br}$ &
$10^2f_{\rm br}$
\\
\hline
0.00 & 390.696  & 390.696 & 0.04251 & 1.18753 & 7.1601 & 0.02126 & 1.7900 \\
0.30 & 1469.277 & 435.671 & 0.16473 & 1.18324 & 7.1409 & 0.02044 & 1.7271 \\
0.60 & 1183.030 & 385.202 & 0.13714 & 1.16954 & 7.0547 & 0.02031 & 1.7363 \\
0.93 & 1212.097 & 356.445 & 0.14595 & 1.12625 & 6.8701 & 0.02008 & 1.7825 \\
\end{tabular*}
\end{ruledtabular}
\end{center}
\end{widetext}

\subsection{Observer-projection dependence}
\label{app:inclination_scan}

The third scan varies the observer inclination while keeping $a/M=0.93$ and
$\beta M^2=10^{-4}$ fixed.  The background geometry and the two effective
optical metrics are unchanged; only the way in which the observer cuts and
projects the critical families changes.  Table~\ref{tab:app_inclination_scan_geometric}
shows that the splitting is not tied to a special observer orientation.  The
projection changes with $\theta_o$, but the separation between $\Gamma_+$ and
$\Gamma_-$ remains.  Taken together, the three scans show that the signal is
opened by the constitutive response, reorganized by rotation, and preserved
under changes of local projection inside the perturbative domain considered.

\begin{widetext}

\begin{center}
\refstepcounter{table}\label{tab:app_inclination_scan_geometric}
\begin{minipage}{0.96\textwidth}
\footnotesize
\textbf{TABLE~\thetable.} Scan in the observer inclination.  We fix $a/M=0.93$
and $\beta M^2=10^{-4}$.  Only the local projection is changed.  All rows
satisfy $\epsilon_{\rm pert}\leq0.05$.
\end{minipage}
\vspace{2pt}

\footnotesize
\begin{ruledtabular}
\begin{tabular*}{0.96\textwidth}{@{\extracolsep{\fill}}cccccccc}
$\theta_o({}^\circ)$ &
$10^4\Delta_{\rm cel}^{\rm max}$ &
$10^4\langle\Delta_{\rm cel}\rangle$ &
$\Delta_{\rm scr}^{\rm max}$ &
$d_0$ &
$10^2\delta d_{\rm br}$ &
$w_{\rm br}$ &
$10^2f_{\rm br}$
\\
\hline
17 & 1281.168 & 390.542 & 0.14167 & 1.08968 & 7.3640 & 0.01959 & 1.7981 \\
50 & 1180.365 & 350.516 & 0.13800 & 1.10739 & 7.0867 & 0.01987 & 1.7943 \\
60 & 1157.753 & 352.215 & 0.13707 & 1.11439 & 6.9976 & 0.01994 & 1.7896 \\
85 & 1212.097 & 356.445 & 0.14595 & 1.12625 & 6.8701 & 0.02008 & 1.7825 \\
\end{tabular*}
\end{ruledtabular}
\end{center}

\endgroup
\end{widetext}

\bibliography{biblio_v1}

\end{document}